\documentclass[%
 reprint,
superscriptaddress,
 amsmath,amssymb,
aps,
pre,
floatfix,
]{revtex4-1}

\usepackage[utf8]{inputenc}
\usepackage{graphicx}% Include figure files
\usepackage{dcolumn}% Align table columns on decimal point
\usepackage{bm}% bold math
\usepackage[colorlinks=true]{hyperref}% add hypertext capabilities
\usepackage{bbm}
\usepackage{subfig}
\usepackage{float}
\usepackage{xcolor}

\begin{document}

\title{The Sherrington-Kirkpatrick model for spin glasses: A new approach for the solution}% Force line breaks with \\

\author{C. D. Rodr\'iguez-Camargo}
 \email{cdrodriguezc@unal.edu.co}%Lines break automatically or can be forced with \\
\affiliation{%
Centro de Estudios Industriales y Log\'isticos para la productividad (CEIL, MD)\\
Programa de Ingenier\'ia Industrial\\
Corporaci\'on Universitaria Minuto de Dios, Bogot\'a AA 111021, Colombia
}%
\affiliation{%
 Programa de Investigaci\'on sobre Adquisici\'on y An\'alisis de Se\~nales (PAAS-UN)\\ Universidad Nacional de Colombia, Bogot\'a AA 055051, Colombia
 }%
 \author{E. A. Mojica-Nava}
\email{eamojican@unal.edu.co}
\affiliation{%
Departamento de Ingenier\'ia El\'ectrica y Electr\'onica \\ 
Facultad de Ingenier\'ia \\ 
Universidad Nacional de Colombia, Bogot\'a AA 055051, Colombia
}%
\affiliation{%
 Programa de Investigaci\'on sobre Adquisici\'on y An\'alisis de Se\~nales (PAAS-UN)\\ Universidad Nacional de Colombia, Bogot\'a AA 055051, Colombia
 }%
\author{N. F. Svaiter}
\email{nfuxsvai@cbpf.br}
\affiliation{%
Centro Brasileiro de Pesquisas F\'isicas\\
Rua Dr. Xavier Sigaud, 150, 22290-180, Rio de Janeiro, RJ, Brasil
}%

\begin{abstract}
We discuss the Sherrington-Kirkpatrick mean-field version of a spin glass within the distributional zeta-function method (DZFM). In the DZFM, since the dominant contribution to the average free energy is written as a series of moments of the partition function of the model, the spin-glass multivalley structure is obtained. Also, an exact expression for the saddle points corresponding to each valley and a global critical temperature showing the existence of many stables or at least metastables equilibrium states is presented. Near the critical point we obtain analytical expressions of the order parameters that are in agreement with phenomenological results. We evaluate the linear and nonlinear susceptibility and we find the expected singular behavior at the spin-glass critical temperature. Furthermore, we obtain a positive definite expression for the entropy and we show that ground-state entropy tends to zero as the temperature goes to zero. We show that our solution is stable for each term in the expansion. Finally, we analyze the behavior of the overlap distribution, where we find a general expression for each moment of the partition function.

\end{abstract}

\maketitle

\section{Introduction}

In statistical mechanics, we define critical points as the points where the thermodynamic free energy or its derivative are singular~\cite{galla1, andelman1, nagle1, ach1}. The simplest non-trivial model that presents a phase transition, from paramagnetic to ferromagnetic phase, is the Ising model~\cite{ising1, onsager1}. In order to describe physical systems with antiferromagnetic and ferromagnetic interactions, the concept of spin glass was introduced~\cite{edwards1, griff1}. One of the main characteristics of such systems is that the order parameter at low temperatures has a random spatial structure~\cite{virasoro1, natterman1, dotsenko1, dominicis1}. Another special characteristic is the multivalley structure in the free-energy landscape~\cite{binder1}. To achieve a reasonable description of the rich topography of the energy surface, it has been proposed a wide set of methods such as dynamical methods~\cite{cugliandolo1, cugliandolo2}, replica methods, and others.

Furthermore, by using density functional theory~\cite{yusuf1}, mode coupling theory~\cite{gotze1}, and purely thermodynamical method the metastable states in the equilibrium free-energy landscape in any glassy system are investigated. It is shown that near a critical point, an exponential number of spin-glass-like metastables states appear in the free-energy landscape leading to a complete dynamical freezing of the system~\cite{monasson1}.

The Edwards and Anderson (EA) Hamiltonian of spin-glasses is defined as
\begin{equation}
H=-\sum _{\left\langle i,j\right\rangle} J_{ij}S_{i}S_{j},
\end{equation} 
where $\left\langle i,j\right\rangle$ represents nearest neighbours of a lattice and $J_{ij}$ satisfies some probability distribution $P(J_{ij})$, as for example, a Gaussian probability distribution $P(J_{ij})\sim \exp [-(J_{ij}-J_{0})/2J^{2}]$, being $J_{0}$ is the mean and variance $J^{2}$.

The infinite range version of the model, was solved in~\cite{sk1}, where the mean field approximation is exact. Using the ansatz of replica symmetry a problem of negative entropy at low temperatures appears, see for example~\cite{nishimori1}. Almeida and Thouless showed that the stability condition of the replica-symmetric solution is not satisfied in the region below a line, called the Almeida-Thouless (AT) line~\cite{at1}. 

The solution of the Sherrington-Kirkpatrick (SK) model was obtained by Parisi~\cite{parisi1, parisi2, parisi3, parisi4}. Parisi sugested the replica symmetry breaking (RSB) as a consistent scheme to break the permutational symmetry of fictitious copies of the system introduced by the replica method where $\mathbb{E}[\log Z]=\lim _{n\rightarrow 0}(\mathbb{E}[Z^{n}]-1)/n$. The RSB in disordered spin systems has a physical interpretation related with the emergence of a spin glass phase characterized by many pure states organized in an ultrametric structure~\cite{mezard1, rotondo1, rammal1}. Despite the equations of the RSB-Parisi ansatz were rigorously proven to be exact~\cite{talagrand1}, other methods and solutions have been explored in regimes for $T=0$ or near the critical point~\cite{oppermann1, pankov1, schmid1, oppermann2, crisanti0, crisanti1, crisanti2, parisi11, marinari1}. Extensions of the RSB scheme on random graphs and neural networks are presented in~\cite{agliari0, lupo1, stephan1} and~\cite{amit1, hemmen1, dot1, agliari1}, respectively. However, within all those approaches there are open questions related to the order parameters and spin-glass systems~\cite{sourav1}. 

Recently, it has been proposed an alternative method to average the disorder-dependent free energy in statistical field theory called the distributional zeta function method (DZFM)~\cite{svaiter1}. Within this approach, the dominant contribution to the average free energy is expressed as a series of the integer moments of the partition function of the model~\cite{svaiter2, svaiter3, svaiter4, svaiter5, svaiter6, cdrc1}.

%The DZFM has been used successfully to study the Landau-Ginzburg approach in replica field theory~\cite{svaiter2}, the disordered $\lambda \varphi ^{4}+\rho \varphi ^{6}$ Landau-Ginzburg model~\cite{svaiter3}, disordered Bose-Einstein condensate in hard walls trap~\cite{svaiter4}, multiplicative noise in Euclidean Schwarzschild manifold~\cite{svaiter5}, and more recently for polymers in random media~\cite{svaiter6} and even has been used for constructing performance measures of complex networks~\cite{cdrc1}.  Furthermore, it has been mentioned its potential for entanglement networks in random media~\cite{svaiter7}.

The aim of this paper is to explore the complexity of the free-energy landscape of the SK model using of the DZFM. Since we have an expansion of free energy where all the integer moments of partition function are contributing, we are able to investigate the multi-valley structure each by each minimum. We show that the Parisi results can be recovered using this formalism. We examine the connection between the DZFM and the phenomenological characterization of the spin-glass phase. We obtain an order parameter $q_{k}$ and $m_{k}$ for each moment of the partition function. Furthermore, we analyze the low temperature regime and the behaviour near the critical point where we obtain an expression of the critical temperature for each minimum in the series representation of the averaged free energy. Afterward, we examine the local magnetization in the regime $m_{k}>0$ and $J_{0}>0$. We find the critical temperatures where the magnetization is non null and we extract the behavior of the linear susceptibility $\chi _{0}$ and the non linear susceptibility $\chi _{2}$. Keeping terms of $O(m_{k}^{2})$ and $O(q_{k}m_{k})$ we show that our solutions are compatible with the results that have been obtained by phenomenological models and experiments. Finally, we analyze the stability due to the order parameter, and show that our solution is stable for each term in the expansion. We also study the overlap distributions and show a different statistics with respect the induced ultrametricity in the RSB.

The organization of the paper is as follows: In Section \ref{secskmod} we review the SK model, present the DZFM approach to this model, and examine our solution proposed to the order parameters, beside a regularization procedure. Furthermore, we also obtain the ground-state entropy and we show that tends to zero as $T\rightarrow 0$. In Section \ref{propertiesofqk} we present our main result. The order parameters of the SK model derived from the DZFM are presented and we explore its behavior for low temperatures and near the critical point. We perform the calculus of the magnetization (main and local) and the susceptibilities. We study the properties of the stability. Conclusions are presented in Section \ref{conclusions}. In the appendix \ref{replicamethodapp} we review the replica method in order to compare with own. In the appendix \ref{dzfmapp1} we review the derivation of the configurational average free energy from the distributional zeta function. Finally, in the appendix \ref{overlapapp1} we study the overlap distribution.

\section{DZFM approach to Sherrington-Kirkpatrick model}
\label{secskmod}

The Sherrington-Kirkpatrick (SK) is defined by the Hamiltonian
\begin{equation}
H=-\sum_{i<j}J_{ij}S_{i}S_{j}-h\sum _{i}S_{i} .
\end{equation}

The first sum on the right hand side runs over all distinct pairs of spins, $N(N-1)/2$ of them. The interaction $J_{ij}$ is a quenched variable with the Gaussian distribution function
\begin{equation}
P(J_{ij})=\frac{1}{J}\sqrt{\frac{N}{2\pi}}\exp \left\lbrace -\frac{N}{2J^{2}}\left( J_{ij}-\frac{J_{0}}{N}\right) ^{2} \right\rbrace .
\label{distributionj}
\end{equation}

The mean and variance of this distribution are both proportional to $1/N$
\begin{equation}
\mathbb{E}[J_{ij}]=\frac{J_{0}}{N}, \quad \mathbb{E}[(\Delta J_{ij})^{2}]=\frac{J^{2}}{N}.
\end{equation}

The partition function of this system for a given configuration of $J$ yields
\begin{equation}
Z_{J}=\text{Tr}\exp (-\beta H)=\text{Tr} \exp \left( \beta \sum _{i<j}J_{ij}S_{i}S_{j}+\beta h\sum _{i}S_{i} \right) .
\end{equation}

With this definition, we have that the free energy per spin is given by
\begin{equation}
f_{N ,J}=\frac{1}{\beta N}\log Z _{J} .
\label{freesample}
\end{equation}

The configurational averaged free energy is 

\begin{equation}
f_{N}=\mathbb{E}[f_{N ,J}] ,
\end{equation}
where $\mathbb{E}[\cdots]$ denotes the configurational average over the distribution (\ref{distributionj}). In the thermodynamic limit we have
\begin{equation}
f=-\lim _{N \rightarrow \infty}\frac{1}{\beta N} \int \left( \prod _{i<j}d(J_{ij})P(J_{ij})\right) \log Z(J_{ij}) .
\label{averagefree111}
\end{equation} 

When the volume of the system tends to infinity, thanks to the self-averaging property, the free energy of the single sample (\ref{freesample}) (specific realization of the disorder) is given by the average over the disorder of the J-dependent free energy, namely we have,
\begin{equation}
\lim _{N \rightarrow \infty} f_{N ,J} = f = \lim _{N \rightarrow \infty} \mathbb{E}[f_{N ,J}] .
\end{equation}

To calculate the average of the free energy we use the DZFM. The main characteristic of the method is that all integer moments of the partition function contribute in this series representation. We use the definition of the distributional zeta function $\Phi (s)$ inspired by the spectral zeta function given by
\begin{equation}
\Phi (s)=\int d[J_{ij}]P(J_{ij})\frac{1}{Z(J_{ij})^{s}}
\label{distzeta1}
\end{equation}
for $s\in \mathbb{C}$, where this function is defined in the region where the above integral converges. The average free energy can be written as
\begin{equation}
f=\lim _{N \rightarrow \infty}\frac{1}{\beta N}\left. \frac{d}{ds}\Phi (s)\right| _{s=0^{+}},\quad \text{Re}(s)\geq 0 ,
\end{equation}
where (\ref{distzeta1}) is well defined. After some algebraic steps, the average free energy can be represented by
\begin{equation}
f=\lim _{N \rightarrow \infty}\frac{1}{\beta N}\left\lbrace \sum _{k=1}^{\infty}\frac{(-1)^{k}a^{k}}{k!k}\mathbb{E}[Z^{k}]+\log a +\gamma _{e} +R(a)\right\rbrace ,
\label{free1}
\end{equation}
where $a$ is a dimensionless parameter, $\gamma _{e}$ is the Euler's constant and the contribution of $R(a)$ can be made as small as desired, taking $a$ large enough; under certain conditions $R(a)$ is bounded as $|R(a)|\leq (Z_{c}a)^{-1}\exp(-Z_{c}a)$ where $Z_{c}$ is the partition function of a system where $P(J_{ij})=c$, with $c \in \mathbb{R}$ being a constant (see Appendix \ref{dzfmapp1} for further details). On the other hand the configurational average of the $k$-th power of the partition function $\mathbb{E}[Z^{k}]$ is
\begin{eqnarray}
\mathbb{E}[Z^{k}]=\int \left( \prod _{i<j}d(J_{ij})P(J_{ij})\right) Z^{k} ,
\end{eqnarray}
with
\begin{equation}
Z^{k}=\text{Tr}\exp \left( \beta \sum _{i<j} J_{ij} \sum _{\alpha = 1}^{k} S_{i}^{\alpha}S_{j}^{\alpha} + \beta h \sum _{i=1}^{N}\sum _{\alpha =1}^{k}S_{i}^{\alpha} \right)
\end{equation}
where $\alpha \in \mathbb{N}$. Since we are using Greek letters for summed moment indices, let us write a subindex $k$ to stand that for each $k$, the moment indices are running from $1$ to $k$. Performing the integral over $J_{ij}$ and rewriting the sums over $i<j$ and moment index, we have, for sufficiently large $N$,
\begin{widetext}
\begin{equation}
\mathbb{E}[Z^{k}]=\exp \left( \frac{N\beta ^{2}J^{2}k}{4}\right) \text{Tr}\exp \left\lbrace \frac{\beta ^{2}J^{2}}{2N}\sum _{\alpha _{k} < \gamma _{k}}\left( \sum _{i}S_{i}^{\alpha _{k}}S_{i}^{\gamma _{k}}\right) ^{2}+\frac{\beta J_{0}}{2N}\sum _{\alpha _{k}}\left( \sum _{i}S_{i}^{\alpha _{k}}\right) ^{2}+\beta h \sum _{i}\sum _{\alpha _{k}}S_{i}^{\alpha _{k}}\right\rbrace .
\label{epart1}
\end{equation}

The trace over $S_{i}^{\alpha _{k}}$ can be carried out independently each site in (\ref{epart1}) if the quantities in the exponent were linear in the spin variables. Those squared quantities can be linearized by Gaussian integrals with integration variables $q_{\alpha _{k} \gamma _{k}}$ for $(\sum _{i}S_{i}^{\alpha _{k}}S_{i}^{\gamma _{k}})^{2}$ and $m_{\alpha _{k}}$ for $(\sum _{i}S_{i}^{\alpha _{k}})^{2}$ as follows
\begin{eqnarray}
\mathbb{E}[Z^{k}]=\exp \left( \frac{N\beta ^{2}J^{2}k}{4}\right)\int D[q_{\alpha _{k} \gamma _{k}}]\int D[m_{\alpha _{k}}] \exp \left( -\frac{N\beta ^{2}J^{2}}{2}\sum _{\alpha _{k} < \gamma _{k}}q_{\alpha _{k} \gamma _{k}}^{2}-\frac{N\beta J_{0}}{2}\sum _{\alpha _{k}}m_{\alpha _{k}}^{2}+N\log \text{Tr}\, \text{e} ^{L _{k}}\right) ,
\label{epart2}
\end{eqnarray}
where
\begin{equation}
D[q_{\alpha _{k} \gamma _{k}}] = \prod _{\alpha _{k} < \gamma _{k}}dq_{\alpha _{k} \gamma _{k}},\quad D[m_{\alpha _{k}}]=\prod _{\alpha _{k}}dm_{\alpha _{k}} . 
\end{equation}

In the equation (\ref{epart2}) 
\begin{equation}
\exp (N\log \text{Tr}\, \text{e}^{L _{k}})=\left\lbrace \text{Tr}\exp \left( \beta ^{2}J^{2}\sum _{\alpha _{k} < \gamma _{k}}q_{\alpha _{k} \gamma _{k}}S^{\alpha _{k}}S^{\gamma _{k}}+\beta \sum _{\alpha _{k}}(J_{0}m_{\alpha _{k}}+h)S^{\alpha _{k}}\right) \right\rbrace ^{N} ,
\end{equation}
and
\begin{equation}
L _{k}= \beta ^{2}J^{2}\sum _{\alpha _{k} < \gamma _{k}}q_{\alpha _{k} \gamma _{k}}S^{\alpha _{k}}S^{\gamma _{k}}+\beta \sum _{\alpha _{k}}(J_{0}m_{\alpha _{k}}+h)S^{\alpha _{k}} .
\end{equation}

Since the exponent of the above integrand is proportional to $N$, it is possible to evaluate the integral by steepest descent such that (\ref{epart2}) yields
\begin{equation}
\mathbb{E}[Z^{k}]\approx \exp \left( -\frac{N\beta ^{2}J^{2}}{2}\sum _{\alpha _{k} < \gamma _{k}}q_{\alpha _{k}  \gamma _{k}}^{2}-\frac{N\beta J_{0}}{2}\sum _{\alpha _{k}}m_{\alpha _{k}}^{2}+N\log \text{Tr}\, \text{e} ^{L _{k}} +\frac{N}{4}\beta ^{2}J^{2}k \right) ,
\label{epartsteepest}
\end{equation}
\end{widetext}
where the $q_{\alpha _{k}  \gamma _{k}}$ are the saddle points and can be defined by
\begin{equation}
q_{\alpha _{k}  \gamma _{k}} = \frac{1}{N}\sum _{i=1}^{N}S_{i}^{\alpha _{k}}S_{i}^{\gamma _{k}}
\end{equation} 
and $q_{\alpha _{k}  \gamma _{k}} \in [0, 1]$. From (\ref{epartsteepest}), the extremalization condition $\delta (\beta f)/\delta q_{\alpha _{k}\gamma _{k}} = 0$ gives us an expression for the order parameter in terms of the average of spins ensemble
\begin{equation}
q_{\alpha _{k}\gamma _{k}} = \frac{\text{Tr}\left[ S^{\alpha _{k}}S^{\gamma _{k}}\text{e}^{L _{k}} \right]}{\text{Tr}\left[\text{e}^{L _{k}} \right]}= \langle \langle S^{\alpha _{k}}S^{\gamma _{k}} \rangle \rangle ,
\label{qalphagamma1}
\end{equation}
where the $\langle \langle \cdots \rangle \rangle$ denotes the average with respect the weight $\text{e}^{L_{k}}$. While the condition $\delta (\beta f)/\delta m_{\alpha _{k}} = 0$ defines a local magnetization as
\begin{equation}
m_{\alpha _{k}}= \langle \langle S^{\alpha _{k}} \rangle \rangle .
\label{malphak1}
\end{equation} 

With these quantities we can express the total magnetization as
\begin{equation}
m=\lim _{N\rightarrow \infty} \sum _{k=1}^{\infty}\frac{(-1)^{k+1}a^{k}}{k!k}\mathbb{E}[Z^{k}] \sum _{\alpha _{k}}m_{\alpha _{k}},
\label{magm1}
\end{equation}
and the linear susceptibility $\chi _{0}$ as
\begin{equation}
\frac{\chi _{0}}{\beta}= \lim _{N\rightarrow \infty} \sum _{k=1}^{\infty}\frac{(-1)^{k+1}a^{k} (N-1)}{k!k}\mathbb{E}[Z^{k}] \left[ \sum _{\alpha _{k}}m_{\alpha _{k}} \right] ^{2}-q ,
\end{equation}
where we define the \emph{weighted} spin-glass order parameter as
\begin{equation}
q =  \lim _{N\rightarrow \infty} \sum _{k=1}^{\infty}\frac{(-1)^{k}a^{k}}{k!k} \mathbb{E}[Z^{k}] \sum _{\alpha _{k}< \gamma _{k}}q_{\alpha _{k}\gamma _{k}} .
\label{spingq1}
\end{equation}

In the replica method, the replica symmetric ansatz is determined by $q_{\alpha _{k}\gamma _{k}} = q_{\text{rsa}}$ and the RSB yields $q_{\alpha _{k}\gamma _{k}} = q(x)$. Here we propose the following solution $q_{\alpha _{k}\gamma _{k}}=q _{k}$ and $m_{\alpha _{k}} = m _{k}$, for the order parameters (\ref{qalphagamma1}) and (\ref{malphak1}), respectively. After taking the limit $N \rightarrow \infty$ for the last terms, and for a sufficiently large $a$, the dominant contribution of (\ref{free1}), yields
\begin{equation}
\beta f = \lim _{N \rightarrow \infty}\sum _{k=1}^{\infty}\frac{(-1)^{k}a^{k}}{N k!k } \text{e}^{kN \varsigma _{k} } ,
\label{freesymm1}
\end{equation}
with
\begin{eqnarray}
\varsigma _{k} = &-&\frac{1}{4}(k-1)\beta ^{2}J^{2}q _{k}^{2}-\frac{1}{2}\beta J_{0}m _{k}^{2}
\nonumber  \\
&+&\frac{1}{k}\log I _{k}+\frac{1}{4}\beta ^{2}J^{2}.
\label{varzetak11}
\end{eqnarray}

In the Equations (\ref{freesymm1}) and (\ref{varzetak11})
\begin{equation}
I _{k}=\int Dz \exp \left[ k\log 2 \text{cosh}(\beta \psi _{k})-\frac{1}{2}k\beta ^{2}J^{2}q_{k} \right]  ,
\end{equation}

being
\begin{equation}
\psi _{k} = J\sqrt{q _{k}}z+J_{0}m _{k}+h ,
\end{equation}
and

\begin{equation}
Dz = \frac{dz}{\sqrt{2\pi}}\exp \left(-\frac{z^{2}}{2}\right) .
\end{equation}

The extremization condition with respect to $q _{k}$, $\delta f / \delta q_{k} = 0$, gives us the following condition,
\begin{eqnarray*}
-\frac{1}{2}k(k-1)\beta ^{2}J^{2}q_{k}+\left[-\frac{1}{2}k\beta ^{2}J^{2} \right.
\nonumber \\
\left. +\frac{\delta}{\delta q_{k}}\log \int Dz \text{cosh}^{k}(\beta \psi _{k}) \right] = 0 .
\end{eqnarray*}

The functional equation for the order parameter $q_{k}$
\begin{equation*}
q_{k}=\frac{1}{(k-1)\beta J \sqrt{q_{k}}}\frac{\int Dz\, z\, \text{cosh}^{k}(\beta \psi _{k}) \text{tanh}(\beta \psi _{k})}{\int Dz\, \text{cosh}^{k}(\beta \psi _{k})}-\frac{1}{k-1} .
\end{equation*}

Partial integration yields
\begin{equation}
q_{k}=\frac{\int Dz\, \text{cosh}^{k}(\beta \psi _{k})\text{tanh}^{2}(\beta \psi _{k})}{\int Dz\, \text{cosh}^{k}(\beta \psi _{k})} .
\label{qzeta}
\end{equation}

On the same way, the condition $\delta f / \delta m _{k} =0$, gives the following relation for the order parameter $m _{k}$
\begin{equation}
m_{k}=\frac{\int Dz\, \text{cosh}^{k}(\beta \psi _{k})\text{tanh}(\beta \psi _{k})}{\int Dz\, \text{cosh}^{k}(\beta \psi _{k})} .
\label{mzeta}
\end{equation}

\begin{figure}[h]
	\includegraphics[width=0.5\textwidth]{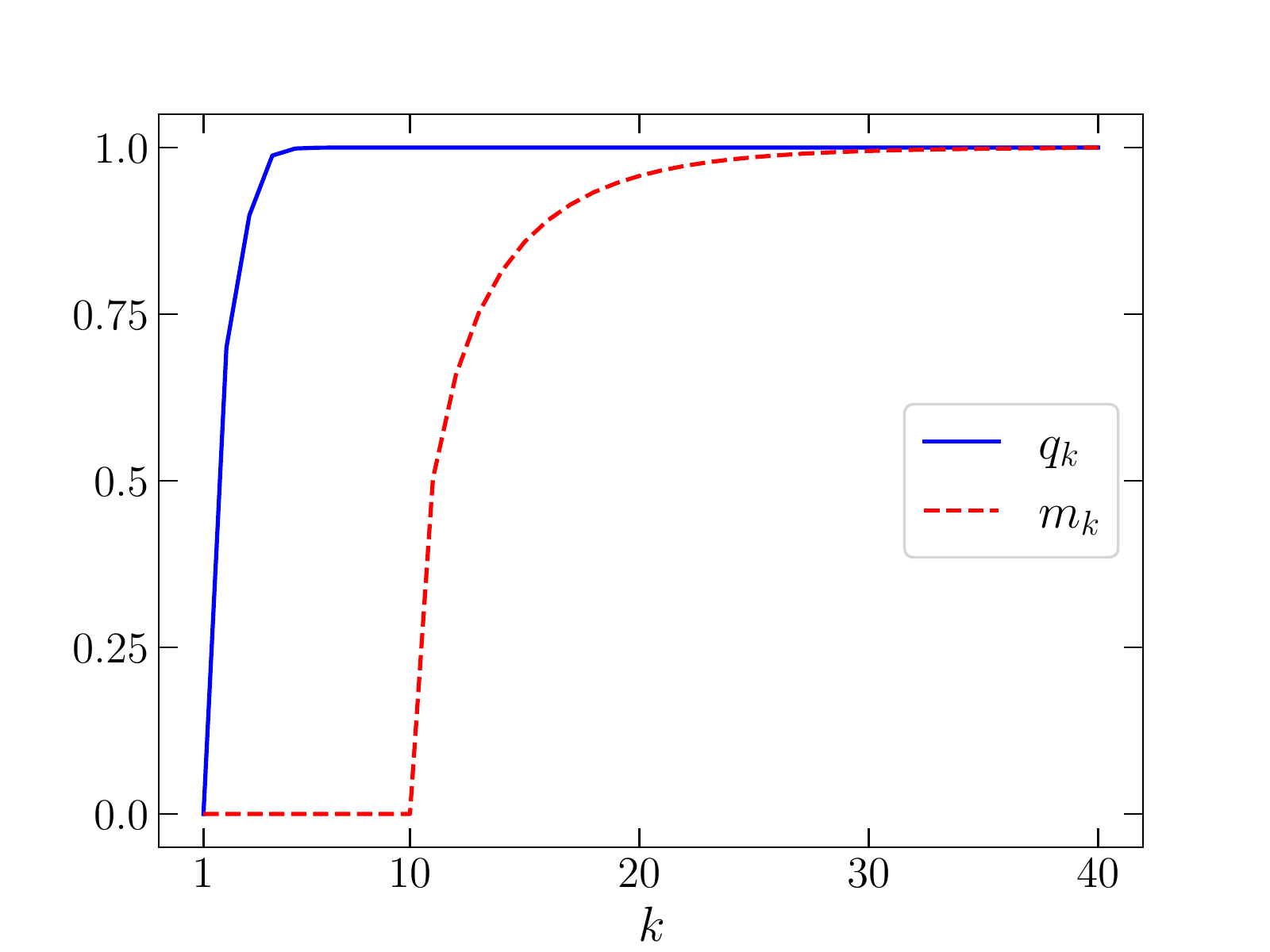}\\
	\caption{Numerical solutions of (\ref{qzeta}) and (\ref{mzeta}) for $J/k_{B}T=1$, $J_{0}/k_{B}T=0.1$, $h=0$ and different $k$. The continuum blue curve describes the behavior of $q_{k}$ and the dashed red curve describes $m_{k}$. We have a transition in these parameters for a given $k$.}
	\label{qandmfork}
 \end{figure}

Notice the structural similarity of Eqs. (\ref{qzeta}) and (\ref{mzeta}) with respect to the results of the replica symmetric ansatz (\ref{qsymm1}) and (\ref{msymm1}), and the 1RSB (\ref{q01rsb}), (\ref{q11rsb}), and (\ref{m1rsb}). In FIG. \ref{qandmfork} we depict the numerical solutions of the equations system (\ref{qzeta}) and (\ref{mzeta}) for, recovering $k_{B}$, $J/k_{B}T=1$, $J_{0}/k_{B}T=0.1$, $h=0$ and different $k$. We may evidence that the index $k$ would be assumed as an order parameter to explore the metastable states in the multi-valley structure of free energy. 

To obtain the ground-state entropy under the DZFM, from the free energy (\ref{freesymm1}), we first derive the low-temperature form of the spin glass parameter $q_{k}$ for $J_{0}=h=0$ and $\beta \rightarrow \infty$ ($T\rightarrow 0$). According to (\ref{qzeta}) we can rewrite that expression as follows
\begin{equation}
q_{k}=1-\frac{\int Dz\, \text{cosh}^{k-2}(\beta J\sqrt{q_{k}}z)}{\int Dz\, \text{cosh}^{k}(\beta J\sqrt{q_{k}}z)}.
\end{equation} 

Thus, for low temperatures,
\begin{eqnarray*}
\int Dz\, \text{cosh}^{k}(\beta J\sqrt{q_{k}}z) \approx  \frac{1}{2^{k}}\int Dz\, \frac{1+k\text{e}^{-2\beta J\sqrt{_{k}q}z}}{\text{e}^{-k\beta J\sqrt{q_{k}}z}} 
\nonumber \\
= \frac{1}{2^{k}}\left[ \text{e}^{k^{2}\beta ^{2} J^{2}q_{k}/2}+k \text{e}^{(k-2)^{2}\beta ^{2} J^{2}q_{k}/2}\right] .
\end{eqnarray*} 

Then, for low temperatures the spin glass parameter $q$ yields
\begin{equation}
q_{k}=1-4\text{e}^{\varpi _{k}(k-1)}\left[ 1-k \text{e}^{\varpi _{k}(k-1)}\right] \left[ 1+(k-2) \text{e}^{\varpi _{k}(k-3)}\right] ,
\label{qlowt}
\end{equation}
being $\varpi _{k}= -2\beta ^{2}J^{2}q _{k}$. After this manipulations, from (\ref{qlowt}) we have that $q_{k}\rightarrow 1$ as $T\rightarrow 0$. Furthermore, by implicit derivation $\partial q_{k}/\partial T \rightarrow 0$ as $T\rightarrow 0$. Within these considerations, the ground-state entropy has the following form

\begin{eqnarray}
S=\lim _{N\rightarrow \infty}\sum _{k=1}^{\infty}(-1)^{k+1}a^{k}2^{kN}s_{k}
\label{entropyzeta}
\end{eqnarray}
where
\begin{equation}
s_{k}=\text{e}^{-\frac{1}{4}Nk^{2}\beta ^{2}J^{2}}\left[ \int Dz\, \text{cosh}^{k}(\beta Jz) \right] ^{N} g_{k}(\beta ,J) ,
\end{equation}
with
\begin{eqnarray}
g_{k}(\beta ,J) = \frac{1}{Nk!k}+\frac{\beta ^{2}J^{2}}{k!}\left( -\frac{k}{2}+(k-1)\iota _{k} (\beta ,J) \right) .
\label{entropygk}
\end{eqnarray}
being
\begin{equation}
 \iota _{k} (\beta ,J)= \frac{\int Dz\, \text{cosh}^{k-2}(\beta Jz)}{ \int Dz\, \text{cosh}^{k}(\beta Jz)} .
\end{equation}

\begin{figure}[h]
	\includegraphics[width=0.5\textwidth]{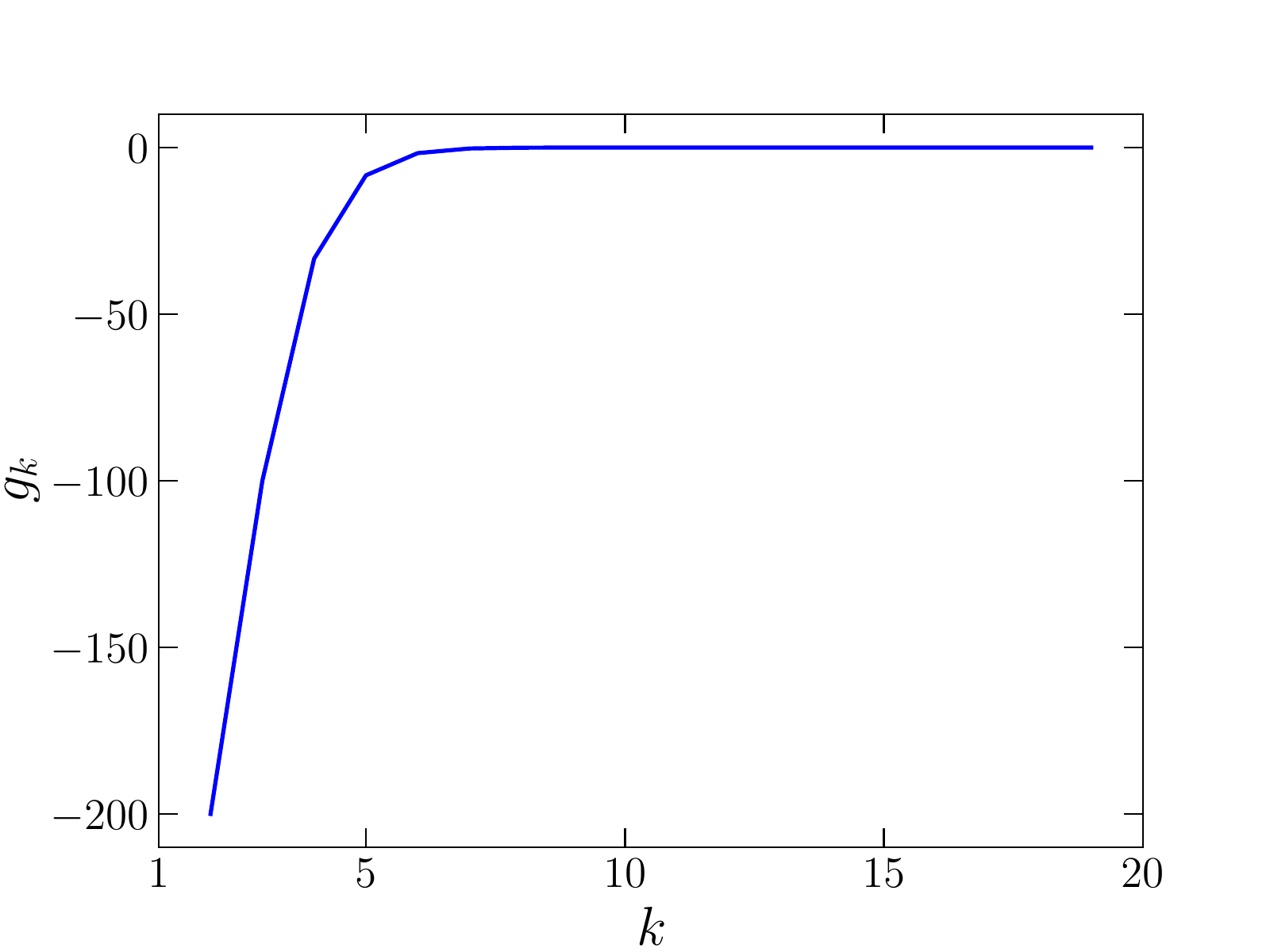}\\
	\caption{Behavior of (\ref{entropygk}) for different $k$, $k_{B}T/J\ll 1$, and large $N$.}
	\label{gkk1}
 \end{figure}

Notice that when we have a symmetric distribution, i.e. $J_{0}=0$, only the values $k=2n$ will contribute. Then, in this case $(-1)^{k+1}$ will be always $-1$. Thus, the entropy (\ref{entropyzeta}) will be positive if $g_{k}(\beta , J)\leq 0$ for all $k$. By simple inspection we have that this condition is accomplished if $\int Dz\, \text{cosh}^{k}(\beta Jz)\geq (2(k-1)/k)\int Dz\, \text{cosh}^{k-2}(\beta Jz)$ which is satisfied for all $k\geq 0$. In the FIG. \ref{gkk1} we depict the behavior of (\ref{entropygk}) where we can evidence that for all $k$ it will be always negative. With this in mind, we have thus that each contribution in (\ref{entropyzeta}) will be positive. Furthermore, we can observe that with the form (\ref{entropyzeta}), $S\rightarrow 0$ as $\beta \rightarrow \infty$ ($T \rightarrow 0$). Here we have obtained with an exact analytical expression the desired limit for the ground-state entropy. In order to obtain high degree analytical expressions we can use the parameter $a$ to regularize our quantities of interest.

\subsection{Regularized quantities}
 
Since $a$ is an arbitrary dimensionless parameter (see Appendix \ref{dzfmapp1}) we can use the quantity $\log a$ to regularize the expression (\ref{freesymm1}), such that we can rewrite (\ref{freesymm1}) as follows
\begin{equation}
f _{r} = \frac{1}{\beta}\sum _{k=1}^{\infty}\frac{(-1)^{k}}{k!}\varsigma _{k} .
\label{freeseries11}
\end{equation}

Notice that (\ref{freeseries11}) and the subsequent quantities are valid on regions where $\varsigma _{k} < 0$. With this form, the total magnetization of the system is
\begin{equation}
m _{r}= \sum _{k=1}^{\infty}\frac{(-1)^{k}}{2k!} \mu _{k},
\label{magnetizationseries1}
\end{equation}
where
\begin{equation}
\mu _{k}=\beta ^{2}J^{2}\eta _{k}[(k-1)q_{k}+1] -2 m_{k}\varrho _{k} 
\end{equation}
being
\begin{equation}
\eta _{k} = (k-2)u_{k}+2m_{k}-km_{k}q_{k} ,
\end{equation}
\begin{equation}
\varrho _{k} = 1-\beta J_{0}[(k-1)q_{k}-km_{k}^{2}+1] 
\end{equation}
and
\begin{equation}
u_{k}=\frac{\int Dz\, \text{cosh}^{k}(\beta \psi)\text{tanh}^{3}(\beta \psi_{k})}{\int Dz\, \text{cosh}^{k}(\beta \psi_{k})} .
\label{uk}
\end{equation}

While the linear susceptibility is given, in analogy with (\ref{magm1}) and (\ref{spingq1}), by 
\begin{equation}
\chi _{0r}= \sum _{k=1}^{\infty}\frac{(-1)^{k}\beta }{2k!}\mu _{k} ^{2}  -q_{r}   ,
\label{susceptibilityseries11}
\end{equation}
where the spin-glass order parameter $q_{r}$ within the aforementioned regularization is defined as
\begin{eqnarray}
q_{r}=\sum _{k=1}^{\infty}\frac{(-1)^{k}\beta}{2k!}\left\lbrace \beta ^{4}J^{4}\eta _{k}^{4}[(k-1)q_{k}+1]^{2} \right.
\nonumber\\
\left. -\beta ^{2}J^{2}b_{k}[(k-1)q_{k}+1] + c_{k} \right\rbrace ,
\label{spingseries111}
\end{eqnarray}
being
\begin{equation}
b_{k}= 4\eta _{k}m_{k}\varrho_{k}+\ell_{k},
\end{equation}
and
\begin{equation}
c_{k}=4m_{k}^{2}\varrho_{k}^{2}-\beta ^{2}J^{2}(k-1)\eta _{k}^{2}-2m_{k}\lambda_{k}+\frac{2\varrho _{k}}{\beta J_{0}}(1-\varrho _{k}) .
\end{equation}

In the above equation
\begin{equation}
\ell _{k} = (k-2)\upsilon _{k}+(2-kq_{k})\frac{1-\varrho _{k}}{J_{0}}-km_{k}\eta _{k} ,
\end{equation}
\begin{equation}
\lambda _{k} = (k-1)\beta J_{0}\eta _{k}-2km_{k}(1-\varrho _{k}) ,
\end{equation}
and
\begin{equation}
\upsilon _{k}=(k-3)r_{k}+3q_{k}-km_{k}u_{k},
\end{equation}
being
\begin{equation}
r_{k}=\frac{\int Dz\, \text{cosh}^{k}(\beta \psi_{k})\text{tanh}^{4}(\beta \psi_{k})}{\int Dz\, \text{cosh}^{k}(\beta \psi_{k})} .
\label{rk}
\end{equation}

In order to explore the advantages of the functional expressions (\ref{qzeta}) and (\ref{mzeta}), and study its implications on the quantities of interest (\ref{freeseries11}), (\ref{magnetizationseries1}), and (\ref{susceptibilityseries11}), in the next section, we shall explore the properties of these new objects near critical points and at the limit $T\rightarrow 0$.
\section{Near the critical point}
\label{propertiesofqk}

The behavior of the solution of the equations of state (\ref{qzeta}) and (\ref{mzeta}) is determined by the parameters $\beta$, $J$ and $J_{0}$. First, we shall focus on the case without external field $h = 0$ and symmetric distribution of $J_{ij}$ ($J_{0}=0$). In this case, we have $\psi _{k}= J\sqrt{q _{k}}z$ such that $\text{tanh} (\beta \psi _{k})$ is an odd function. Then the magnetization vanishes ($m_{k}=0$ for all $k$) and there is no ferromagnetic phase. Afterward we explore the solutions where $J_{0}>0$, $m_{k}>$, and $q_{k}>0$.

\begin{figure}[h]
	\includegraphics[width=0.5\textwidth]{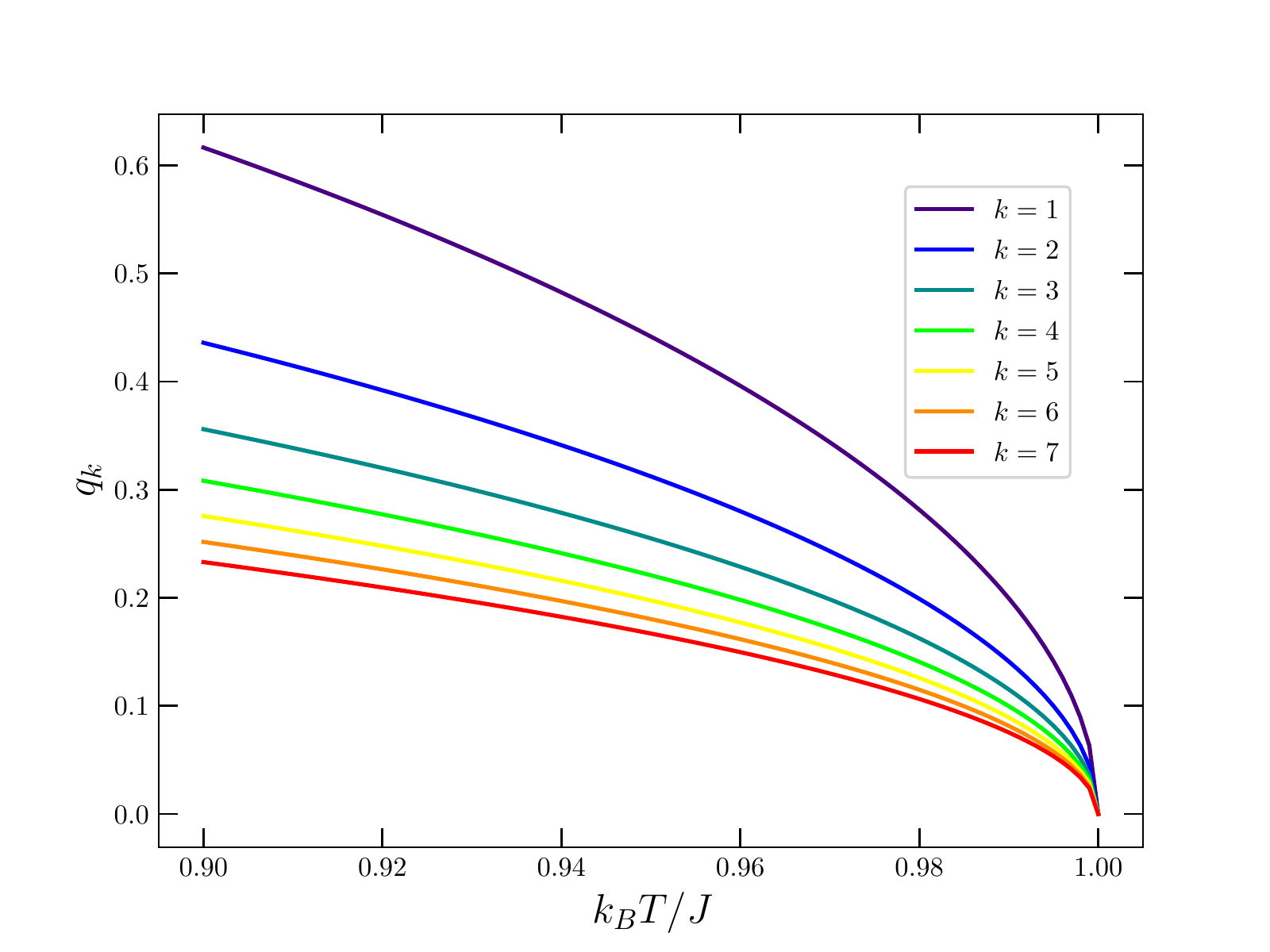}\\
	\caption{Numerical solutions up to $O(\beta ^{2})$ for (\ref{qzeta}) near the critical point for different values of $k$. We can identify the critical temperature at $\beta J = 1$.}
	\label{qkk11}
 \end{figure}

\begin{figure}[h]
	\includegraphics[width=0.5\textwidth]{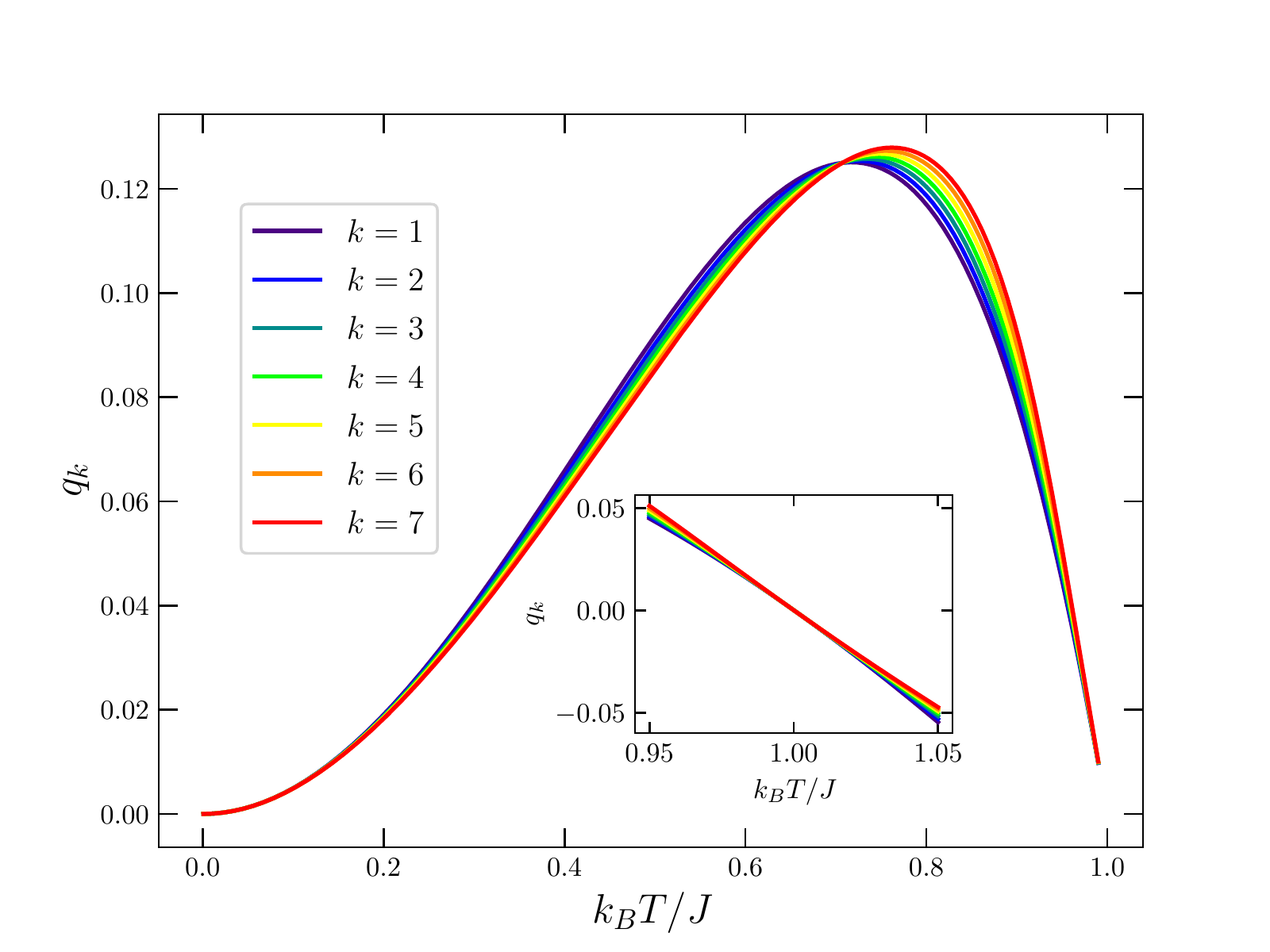}\\
	\caption{Numerical solutions for (\ref{qzeta}), and for the complete expansion. We can identify the critical temperature at $\beta J = 1$. Inset shows the behavior with this expansion near and beyond the point $\beta J = 1$.}
	\label{qkk1}
 \end{figure}

To investigate the properties of the system near the critical point where the spin glass order parameter $q_{k}$ is small, we expand the right hand side of (\ref{qzeta}) and keeping up to $O(q_{k}^{3})$, for $q_{k}\rightarrow 0$, we have $\text{cosh}^{k}(\beta J\sqrt{q _{k}}z)\approx 1+\frac{1}{2}k\beta ^{2}J^{2}q _{k}z^{2}+(-\frac{1}{12}k\beta ^{4}J^{4}z^{4}+\frac{1}{8}\beta ^{4}J^{4}k^{2}z^{4})q_{k}^{2}+O(q _{k}^{3})$, and $\text{tanh}^{2}(\beta J\sqrt{q _{k}}z)\approx \beta ^{2} J^{2}q _{k}z^{2}-\frac{2}{3}\beta ^{4}J^{4}q_{k}^{2}z^{4} + O(q _{k}^{3})$. Replacing the above expansions in (\ref{qzeta}) we obtain the behavior of the numerical solutions for $q_{k}$ depicted in FIG. \ref{qkk11} and FIG. \ref{qkk1} where we can identify the critical temperature given when $\beta ^{2}J^{2}$ becomes one for $J_{0}=0$. In FIG. \ref{qkk11} we show the behavior with a lower degree expansions, near the critical point $\beta J = 1$. In FIG. \ref{qkk1}, we depict the numerical solutions for the complete aforementioned expansions (up to $O(q_{k}^{3})$). As we can observe, $q_{k}\rightarrow \theta$, with $\theta \ll 1$, as we are approaching to the critical point and $k \rightarrow \infty$. Notice that it is a parallel situation to the Parisi scheme where, in the limit of $k \rightarrow 0$ one has $(1/k)\sum _{\alpha \neq \beta}q_{\alpha \beta}^{l} \rightarrow -\int _{0}^{1}q^{l}(x)dx$, and then the order parameter is defined by (\ref{parisifullrsb}). An advantage of this expansion, is that, on the interval $[0, 1]$ we have a similar behavior presented in Ref.~\cite{nishimori1}, i.e., $q(x)\rightarrow 0$ as $x \rightarrow 0$, where the parameter $x$ has a relationship with the parameter $k_{b}T/ J$ (see the complete discussion of the subsection 3.3.2. \emph{Order parameter near the critical point} in Ref.~\cite{nishimori1}). In FIG. \ref{qkk1} the inset shows the behavior with this expansion near and beyond the point $\beta J = 1$.

\begin{figure}[h]
	\includegraphics[width=0.45\textwidth]{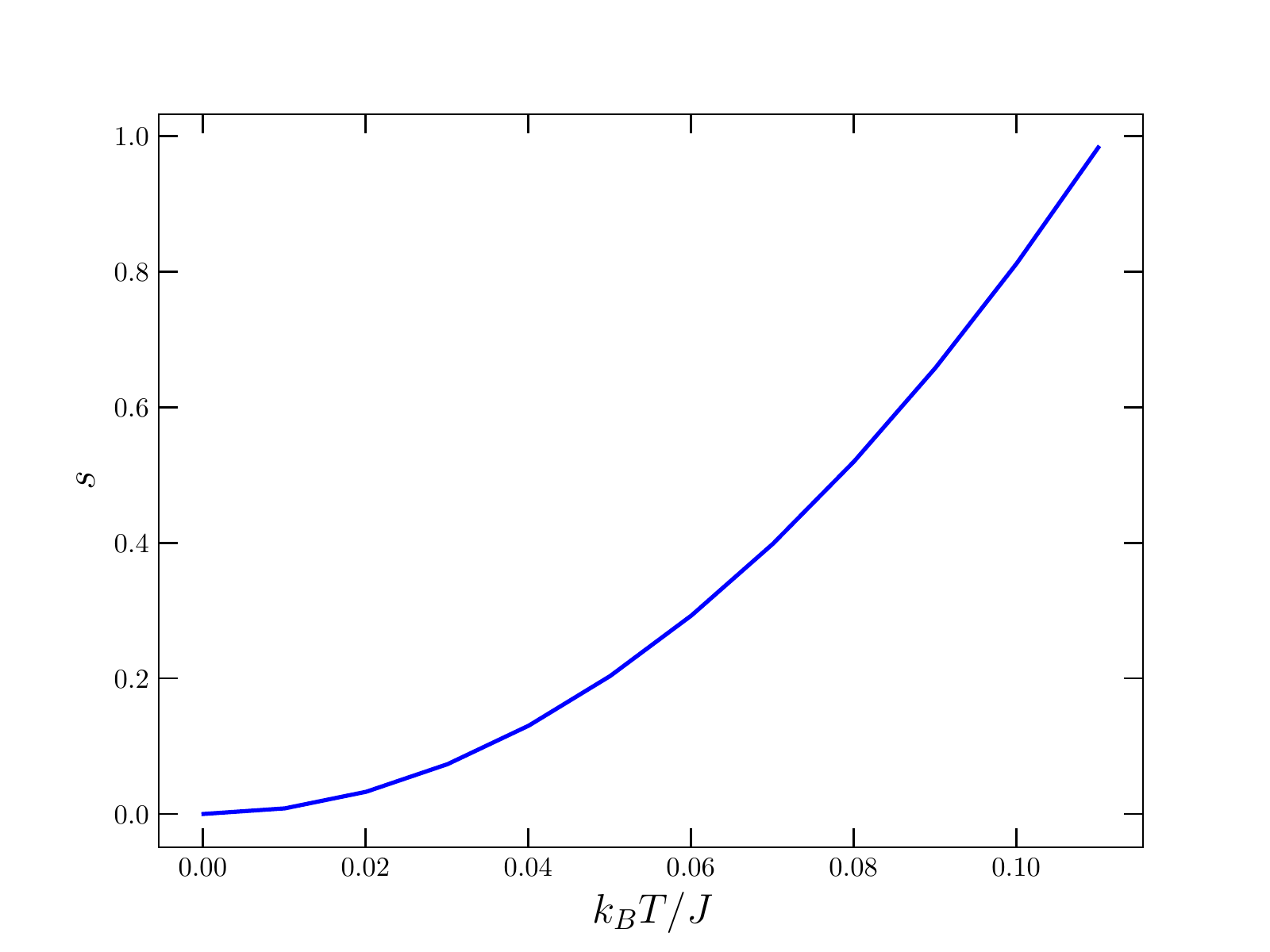}\\
	\caption{Entropy for the numerical solutions for (\ref{qzeta}) and a fixed $J\ll 1$.}
	\label{entropy000}
 \end{figure}

 Furthermore, using this numerical solution for a $J\ll 1$ (inside a region where $\varsigma _{k}<0$ in this regime) we are able to reproduce the aforementioned result for the ground-state entropy when $T\rightarrow 0$. In the Fig. \ref{entropy000} we depict the behavior of this ground-state entropy given by
\begin{equation}
    s=\sum _{k=1}^{\infty}\frac{(-1)^{k+1}}{k!}s_{0k},
\end{equation}
where
\begin{eqnarray}
s_{0k}=& &\beta^{2}J^{2}\left( \frac{1}{4}(k-1)q_{k}^{2}+\frac{1}{2}q_{k}-\frac{1}{4}\right)
\nonumber \\
&+&\frac{1}{2}\beta^{3} J^{2}\frac{\partial q_{k}}{\partial \beta}((k-1)q_{k}+1)
\nonumber \\
&+&\log 2 + \frac{1}{k}\log \int Dz\, \text{cosh}^{k}(\beta J\sqrt{q_{k}}z)
\nonumber \\
&-& \beta ^{2}J^{2}\left( \frac{1}{2}\beta\frac{\partial q_{k}}{\partial \beta}+ q_{k}  \right) ((k-1)q_{k}+1)
\end{eqnarray}

Now, if $J_{0}\ne 0$ we will have four kind of solutions: i) $q_{k}=0$ and $m_{k}=0$; ii) $q_{k}>0$ and $m_{k}=0$; iii) $q_{k}>0$ and $m_{k}>0$; and iv) $q_{k}>0$ and $m_{k}<0$. To explore the regime where $J_{0}>0$, and $m_{k}>0$, we expand (\ref{qzeta}) and (\ref{mzeta}) for small $q_{k}$ and $m_{k}$. If we neglect terms of $O(q_{k}m_{k})$ and keep with $O(q_{k})$ and $O(m_{k}^{2})$, we obtain expressions of the form
\begin{equation}
q_{k}=\frac{2(\beta J_{0}-1)}{k\beta J_{0}},
\label{qkforj0}
\end{equation}
\begin{equation}
m_{k}=\left[\frac{2(\beta J_{0}-\beta ^{2}J^{2})(\beta J_{0}-1)}{k\beta ^{3}J_{0}^{3}}\right]^{1/2} .
\label{mkforj0}
\end{equation}

\begin{figure}[h]
	\includegraphics[width=0.45\textwidth]{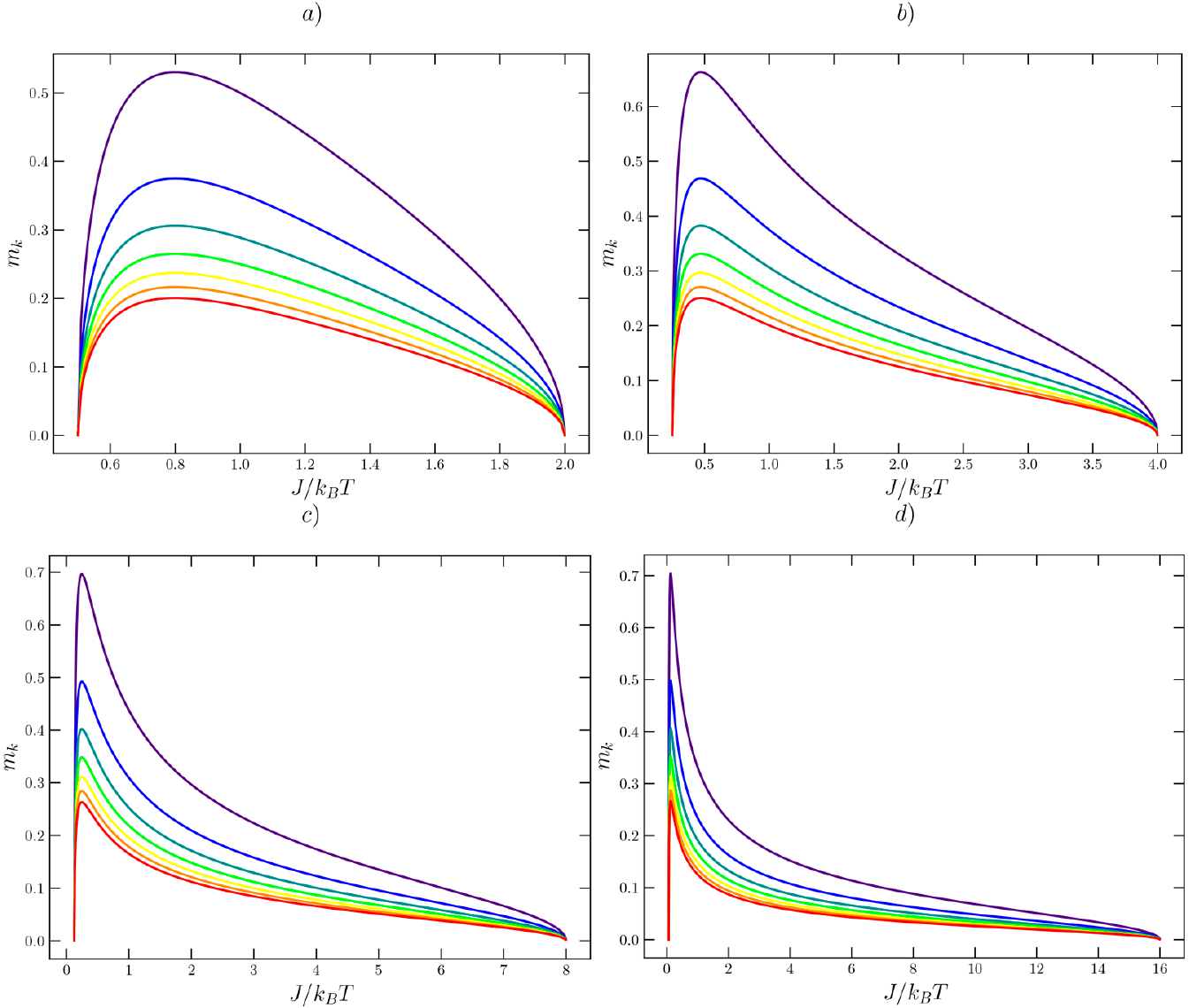}\\
	\caption{Behavior of (\ref{mkforj0}) for (a) $J_{0}/J = 2$, (b) $J_{0}/J = 4$, (c) $J_{0}/J = 8$, (16) $J_{0}/J = 16$. Each curve correspond to $k=1$ (indigo), $k=2$ (blue), $k=3$ (dark cyan), $k=4$ (green), $k=5$ (yellow), $k=6$ (orange), and $k=7$ (red).}
	\label{mk_unidas}
 \end{figure}

Observe that we have, as is expected, the points of transition determined by $\beta J_{0}=1$ or $T_{C}=J_{0}$ and $\beta J^2/J_{0}=1$. In the FIG. \ref{mk_unidas} we depict the behavior of (\ref{mkforj0}) for different temperatures and values of $J_{0}$. Note that the interval where it is well defined is $[J/J_{0},J_{0}/J]$. We can evidence the appearance of a maximum for a given value of $J/k_{B}T$ and $k$. Each curve correspond to different values of $k$. Similar behaviors are obtained experimentally (see Refs.~\cite{binder1, gron1, nirmala1, haldar1, vincent1}).

With these results we are able to recover the typical phase diagram with the regularized order parameters (\ref{magnetizationseries1}) and (\ref{spingseries111}). Furthermore, keep now the terms of $O(q_{k}m_{k})$ to observe other phenomena, the local susceptibility from (\ref{mzeta}) yields
\begin{equation}
\chi _{0k}= -k\beta m_{k}^{2}+(k-1)\beta q_{k}+\beta ,
\label{chi0k}
\end{equation}
while the nonlinear susceptibility reads
\begin{eqnarray}
\chi _{2k}=&-&2k\beta \chi _{0k}^{2}+\beta ^{2}\chi _{0k}[(k-1)(2-kq_{k})-4k^{2}m_{k}]
\nonumber\\
&+&k(k-1)\beta m_{k}[(k-2)u_{k}+m_{k}(2-kq_{k})]
\nonumber\\
&+&(k-1)(k-2)\beta ^{3}[(k-3)r_{k}+3q_{k}-km_{k}u_{k}] .
\nonumber\\
\label{chi2k}
\end{eqnarray}

\begin{figure}[h]
	\includegraphics[width=0.5\textwidth]{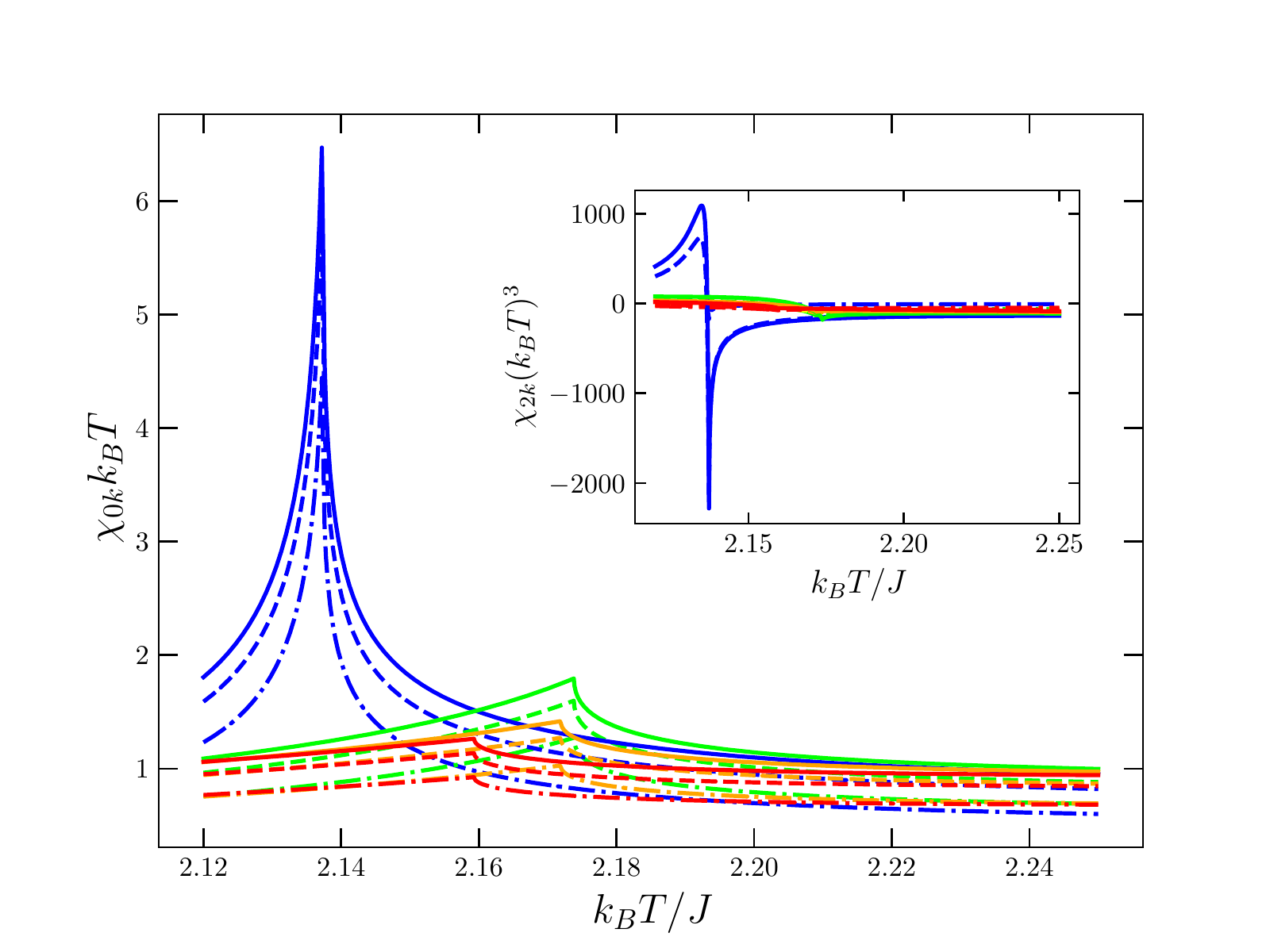}\\
	\caption{Behavior of (\ref{chi0k}) for $J_{0}/J = 3$ (blue), $J_{0}/J = 4$ (lime), $J_{0}/J = 5$ (orange), $J_{0}/J = 6$ (red), and $k=3$ (dot-dashed line), $k=4$ (dashed line), and $k=5$ (continuous line) in function of $k_{B}T/J$. Note the appearance of the discontinuity at the spin glass critical temperature. Inset shows the typical behavior of the nonlinear susceptibility (\ref{chi2k}).}
	\label{chik0111}
 \end{figure}

We depict the behavior of (\ref{chi0k}) in function of $k_{B}T/J$ for  different values of $J_{0}/J$ in the FIG. \ref{chik0111}. Notice that we obtain the expected discontinuity of the linear susceptibility at the spin glass critical temperature extracted from phenomenological models and experimental results~\cite{binder1, suzuki1, chalupa1, suzuki2, bitla1}. Also note that each $k$ (continuous, dot-dashed, and dashed lines) is depicting a similar behavior, as is occurring with the other quantities. Furthermore, the inset shows (\ref{chi2k}). We recover the typical behavior of the nonlinear susceptibility. However, each $k$-contribution is showing a local behavior of each term of the principal expansion. In order to obtain the total critical temperature, for obtaining numerical values to compare with the experimental ones, we have to use the regularized quantities and to find experimental arrays which are in regimes where our regularization procedure is valid. Within our regularized quantities obtained, we can carry out a more complete characterization by obtaining quantities, beside the regularized total magenetization (\ref{magnetizationseries1}), such as, the regularized magnetic susceptibility~\cite{colecole1, levy22}
\begin{equation}
\chi _{r} (\omega)=\chi _{rs}+\frac{\chi _{0r}-\chi _{rs}}{1+(i\omega \tau _{C})^{1-\alpha}}
\end{equation}
being $\tau _{c}$ the characteristic relaxation time while the parameter $\alpha$ ranges from 0 to 1, $\chi _{0r}$ is given by (\ref{susceptibilityseries11}), and $\chi _{rs}$, the regularized adiabatic susceptibility, reads
\begin{equation}
\chi _{rs}= \chi _{0r} - \left( \frac{\partial m _{r}}{\partial T} \right) ^{2} \left( \frac{\partial s _{r}}{\partial T} \right) ^{-1} ,
\end{equation}

which can be used to derive $\chi'_{r} (\omega)$ and $\chi ''_{r}(\omega)$ as follows
\begin{equation}
\chi'_{r} (\omega) = \chi _{rs}+\frac{\chi _{0r}-\chi _{rs}}{1+\omega ^{2}\tau_{C}^{2}},
\end{equation}
and
\begin{equation}
\chi''_{r} (\omega) = \omega \tau _{C}\frac{\chi _{0r}-\chi _{rs}}{1+\omega ^{2}\tau_{C}^{2}} .
\end{equation}

We can obtain too the specific heat
\begin{equation}
C_{f}=-T \frac{\partial ^{2}f_{r}}{\partial T^{2}} ,
\end{equation}

and other quantities that usually are studied on the laboratory~\cite{binder1, suzuki1, chalupa1, suzuki2, bitla1, perovic1, fernandez1}. However, this surpass the scope of this paper and it will be treated in a future work.

\begin{figure*}[htb]
\centering
  \begin{tabular}{@{}cc@{}}
    \includegraphics[width=.34\textwidth]{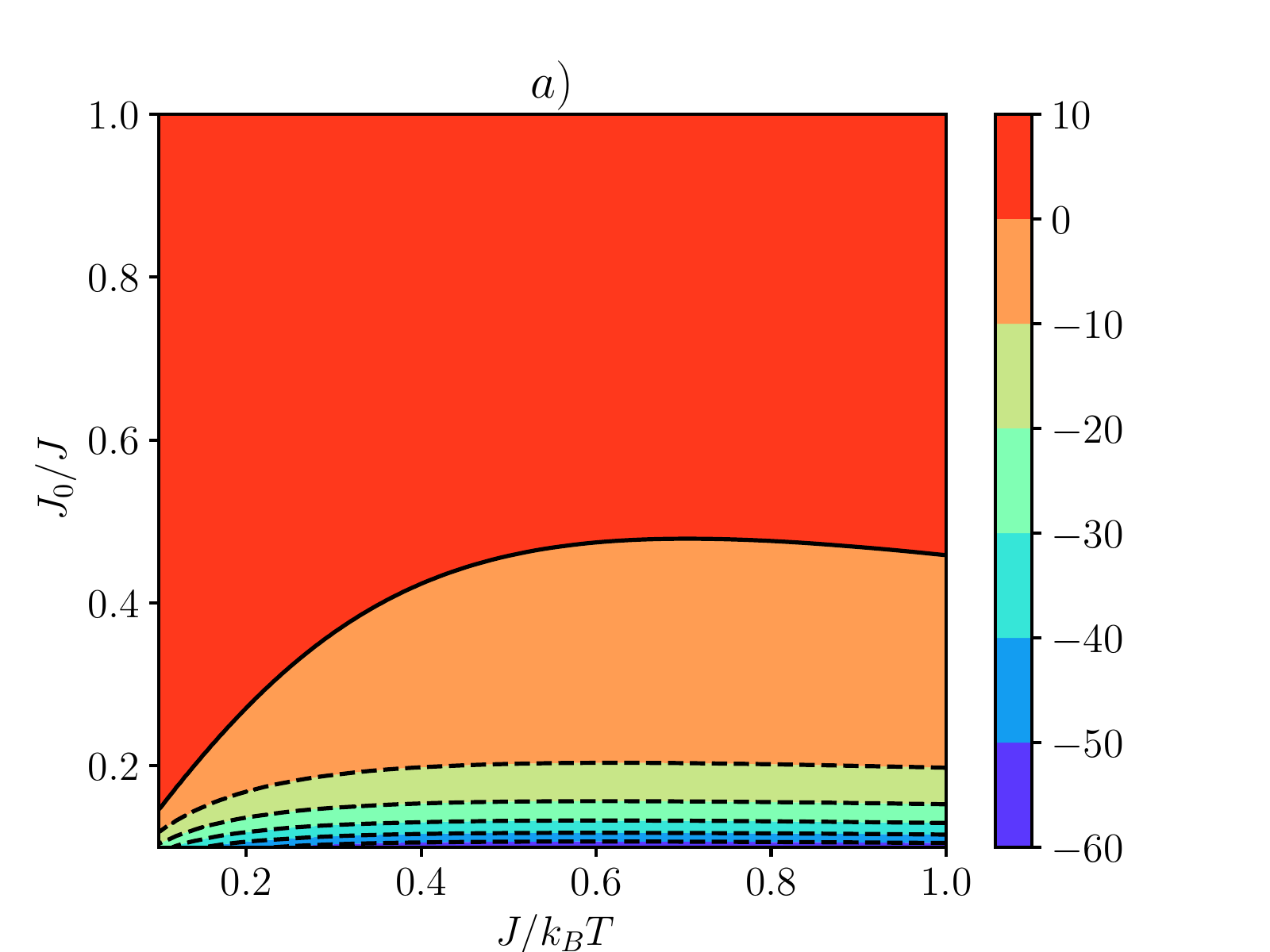} &
    \includegraphics[width=.34\textwidth]{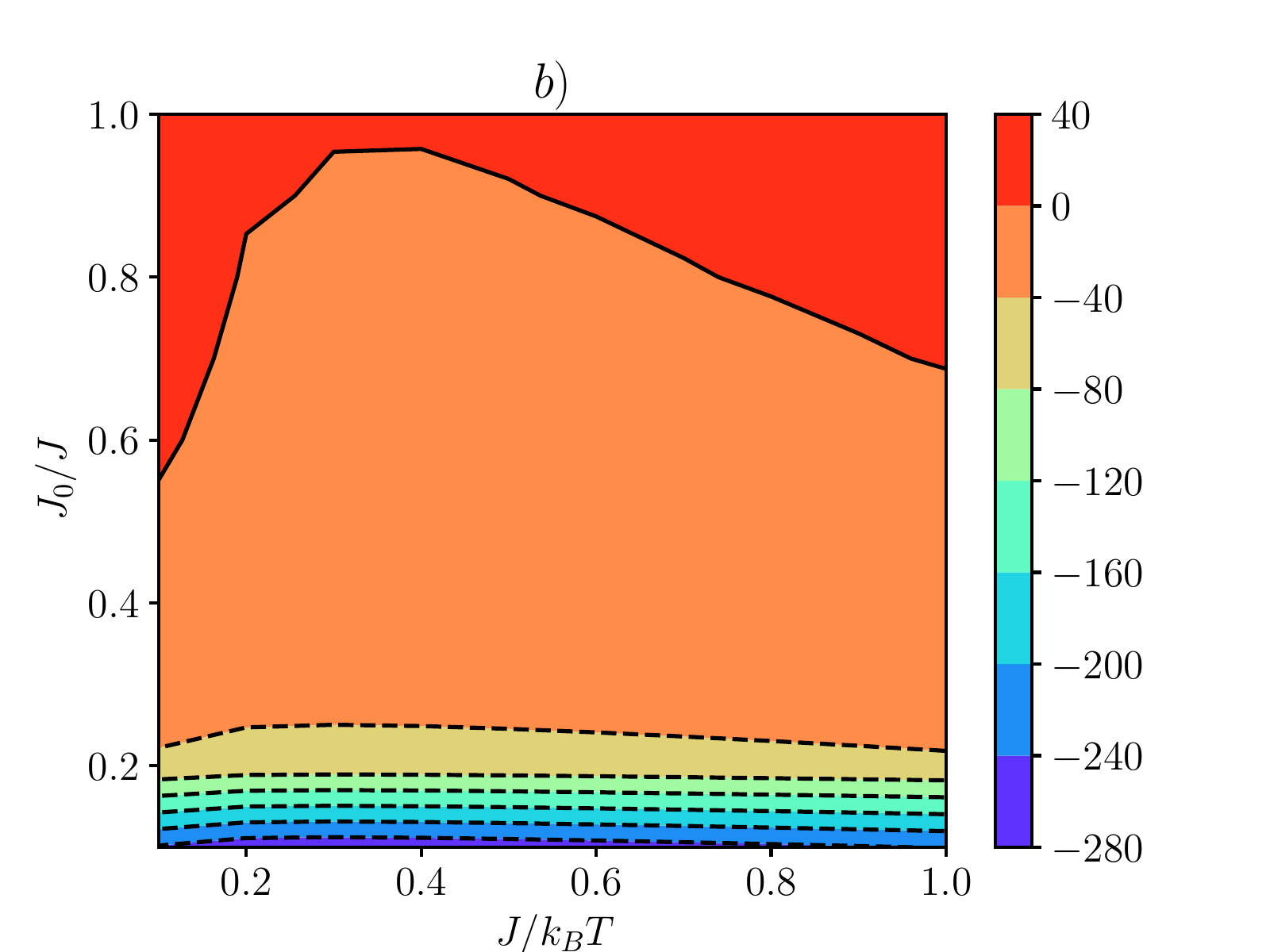} \\
    \includegraphics[width=.34\textwidth]{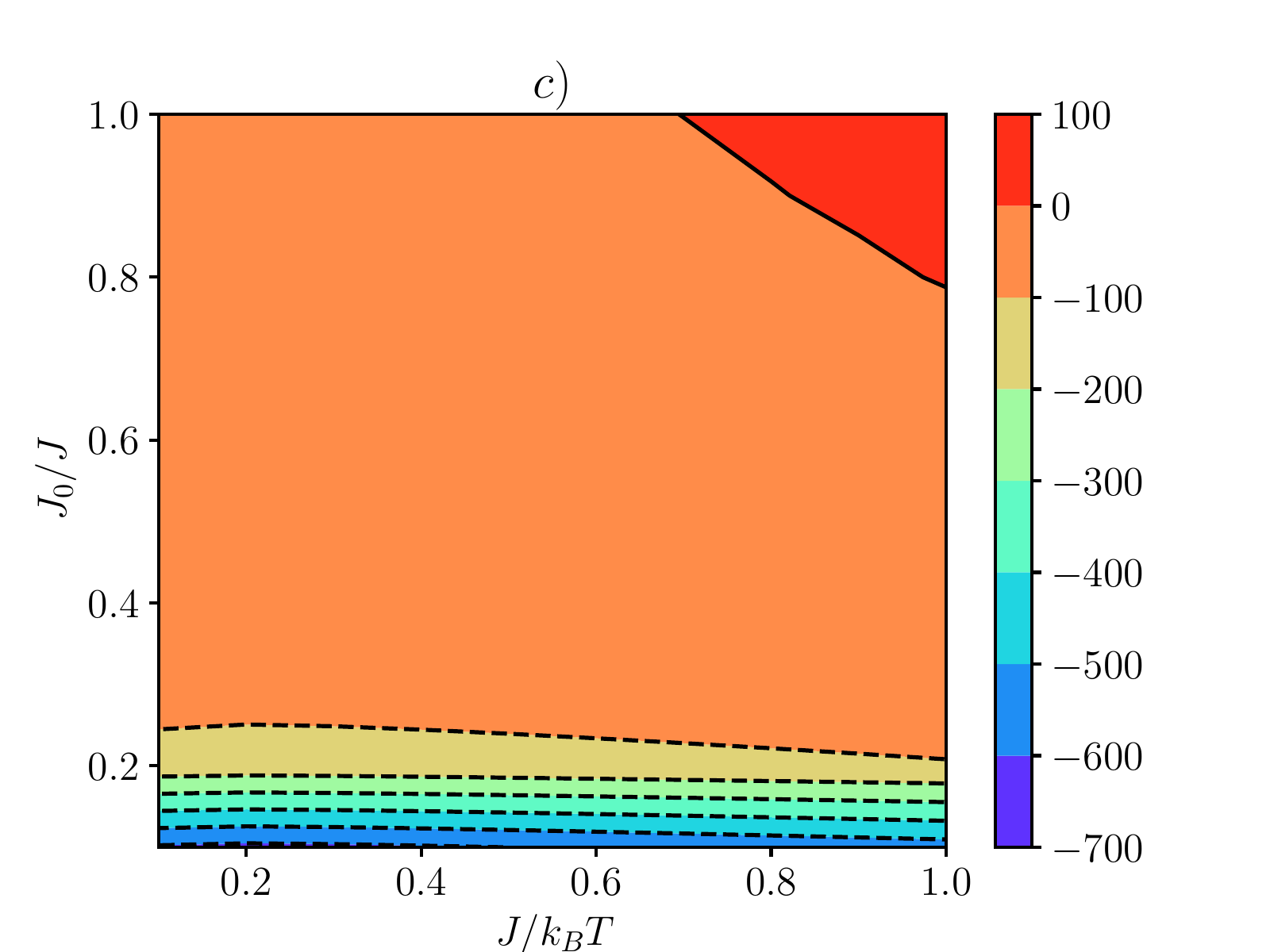} &
    \includegraphics[width=.34\textwidth]{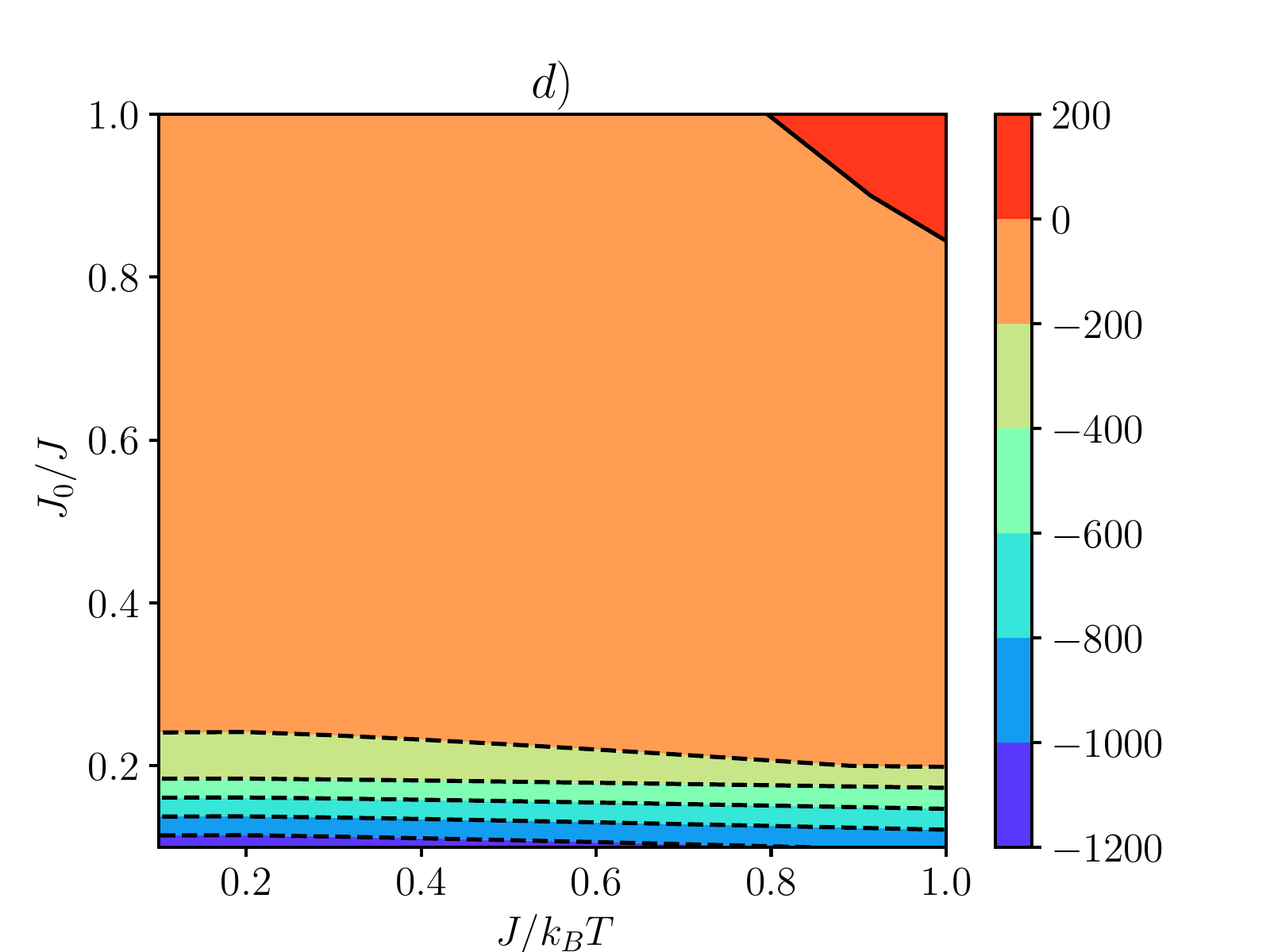} \\
  \end{tabular}
  \caption{Regions of validity of regularized free energy (\ref{freeseries11}) and subsequent regularized quantities for (a) $J=1$, (b) $J=2$, (c) $J=3$, (d) $J=4$.}
  \label{valofzeta}
\end{figure*}

To characterize the aforementioned kind of regimes, we can use the expressions (\ref{mkforj0}) and (\ref{qkforj0}), and explore the regions where (\ref{varzetak11}) is negative. Near the critical point in FIG. \ref{valofzeta} we depict the regions of validity of our regularized quantities. We can evidence that the region is increasing as the disorder becomes stronger (increasing value of $J$). With this result, now we shall study the stability of our solution $q_{\alpha _{k}\gamma _{k}}=q_{k}$.

\subsection{Stability}

For the analysis of the stability, let us explore the Hessian matrix for each term in the series expansion of free energy, for $h=0$. Thus, for each $k$ the Hessian evaluated in $q_{k}=\langle \langle S^{\alpha _{k}}S^{\beta _{k}} \rangle \rangle$ is
\begin{equation}
\frac{\delta ^{2}(\beta f)}{\delta  q_{\alpha _{k}\beta _{k}} \delta  q_{\mu _{k}\nu _{k}}}=\frac{(-1)^{k+1}a^{k}}{k!k}\text{e}^{w_{k}}\frac{\delta ^{2}w_{k}}{\delta  q_{\alpha _{k}\beta _{k}} \delta  q_{\mu _{k}\nu _{k}}}, 
\label{hessian1}
\end{equation}
where
\begin{equation}
w_{k}=\beta ^{2}J^{2} \sum _{\alpha _{k}<\gamma _{k}}q_{\alpha _{k}\gamma _{k}}^{2} S^{\alpha _{k}}S^{\gamma _{k}}+\beta J_{0}\sum _{\alpha _{k}}m_{\alpha _{k}}S^{\alpha _{k}} ,
\end{equation}
and
\begin{eqnarray}
\frac{\delta ^{2}w_{k}}{\delta  q_{\alpha _{k}\beta _{k}} \delta  q_{\mu _{k}\nu _{k}}} =& & \beta ^{2}J^{2}\delta _{(\alpha _{k}\beta _{k}),(\mu _{k}\nu _{k})}
\nonumber \\
&-&\beta ^{4}J^{4}[\langle \langle S^{\alpha _{k}}S^{\beta _{k}}S^{\mu _{k}} S^{\nu _{k}} \rangle \rangle 
\nonumber \\
&-&\langle \langle S^{\alpha _{k}}S^{\beta _{k}} \rangle \rangle \langle \langle S^{\mu _{k}}S^{\nu _{k}} \rangle \rangle] .
\end{eqnarray}

Following the standard procedure~\cite{crisanti1, morone1}, we have three kind of contributions in (\ref{hessian1}):
\begin{equation}
\frac{\delta ^{2}(\beta f)}{\delta  q_{\alpha _{k}\beta _{k}} \delta  q_{\mu _{k}\nu _{k}}}=\begin{cases}
\mathcal{H}_{1}^{(k)}\quad \text{if}\quad (\alpha _{k}\beta _{k})=(\mu _{k}\nu _{k})\\
\mathcal{H}_{2}^{(k)}\quad \text{if}\quad \alpha _{k}=\mu _{k}, \beta _{k}\neq \nu _{k}\\
\mathcal{H}_{2}^{(k)}\quad \text{if}\quad \alpha _{k}\neq \mu _{k}, \beta _{k}= \nu _{k}\\
\mathcal{H}_{3}^{(k)}\quad \text{if}\quad \alpha _{k}\neq \mu _{k}, \beta _{k}\neq \nu _{k}\\
\end{cases} ,
\end{equation}
where each contribution is given by
\begin{equation*}
\mathcal{H}_{1}^{(k)} = \frac{(-1)^{k+1}a^{k}}{k!k}\text{e}^{w_{k}}[\beta ^{2}J^{2}-\beta ^{4}J^{4}(1-q_{k}^{2})],
\end{equation*} 
\begin{equation}
\mathcal{H}_{2}^{(k)} = \frac{(-1)^{k+1}a^{k}}{k!k}\text{e}^{w_{k}}[-\beta ^{4}J^{4}(q_{k}-q_{k}^{2})],
\end{equation} 
\begin{equation*}
\mathcal{H}_{3}^{(k)} = \frac{(-1)^{k+1}a^{k}}{k!k}\text{e}^{w_{k}}[-\beta ^{4}J^{4}(r_{k}-q_{k}^{2})],
\end{equation*} 
with $r_{k}$ defined by (\ref{rk}).

We proceed to study the eigenvalues of (\ref{hessian1}) the subspace spanned by the vector with all equal components is a good eigenspace, called \emph{longitudinal} (scalar) space, which at the same time, can be one can be divided in other two orthogonal eigenspaces called, respectively, the \emph{anomalous} (vectorial) and \emph{replicon} (tensorial).

Let us study the condition of stability for each eigenspace. First, for the longitudinal space, the eigenvalue is given by
\begin{equation}
\lambda _{L}^{(k)}=\mathcal{H}_{1}^{(k)}+2(k-2)\mathcal{H}_{2}^{(k)}+\frac{(k-2)(k-3)}{2}\mathcal{H}_{3}^{(k)} .
\end{equation}

We must divide the study in two cases. The first one, when $k+1=2n$, with $n\in \mathbb{N}$. The second one, $k+1=2n+1$, with $n\in \mathbb{N}$. For the first case, in the spin glass phase, the stability condition is given by $q_{2n-1}>(2n-3)/(n^{2}+(n/2)-1)$, which is valid for all $n>1$. On the other hand, when $k+1=2n+1$, we have $q_{2n}>(n-1)/n(n-1/2)$, which is valid for all $n>1$ too. In the limit $T\rightarrow 0$ we have stability for all $k+1=2n$. Near the critical point $1>\beta ^{2}J^{2}$.

For the anomalous space, the eigenvalue is
\begin{equation}
\lambda _{A}^{(k)}=\mathcal{H}_{1}^{(k)}+(k-4)\mathcal{H}_{2}^{(k)}-(k-3)\mathcal{H}_{3}^{(k)} .
\end{equation}

To go further, we discuss two situations. For $k+1=2n$, with $n\in \mathbb{N}$, in the spin glass phase, the stability condition is given for all $n>2.5$. On the other hand, when $k+1=2n+1$, we have that the solution is stable for all $n>2$. In the limit $T\rightarrow 0$ we have stability for all $k+1=2n$. Near the critical point $1>\beta ^{2}J^{2}$. In the limit $T\rightarrow 0$ we have stability for all $k+1=2n$. Near the critical point $1>\beta ^{2}J^{2}$.

Finally, the eigenvalue of the replicon space yields
\begin{equation}
\lambda _{R}^{(k)}=\mathcal{H}_{1}^{(k)}-2\mathcal{H}_{2}^{(k)}+\mathcal{H}_{3}^{(k)} .
\end{equation}

Which gives us our version for the AT line of stability, that is, for $k+1=2n$, in the spin glass phase
\begin{equation}
\left(\frac{T}{J}\right) ^{2}>\frac{\int Dz\, \text{cosh}^{2n-5}(\beta \psi)}{\int Dz\, \text{cosh}^{2n-1}(\beta \psi)},
\end{equation}
and for $k+1=2n+1$
\begin{equation}
\left(\frac{T}{J}\right) ^{2}<\frac{\int Dz\, \text{cosh}^{2n-4}(\beta \psi)}{\int Dz\, \text{cosh}^{2n}(\beta \psi)}.
\end{equation}

\begin{figure*}[htb]
\centering
  \begin{tabular}{@{}ccc@{}}
    \includegraphics[width=.34\textwidth]{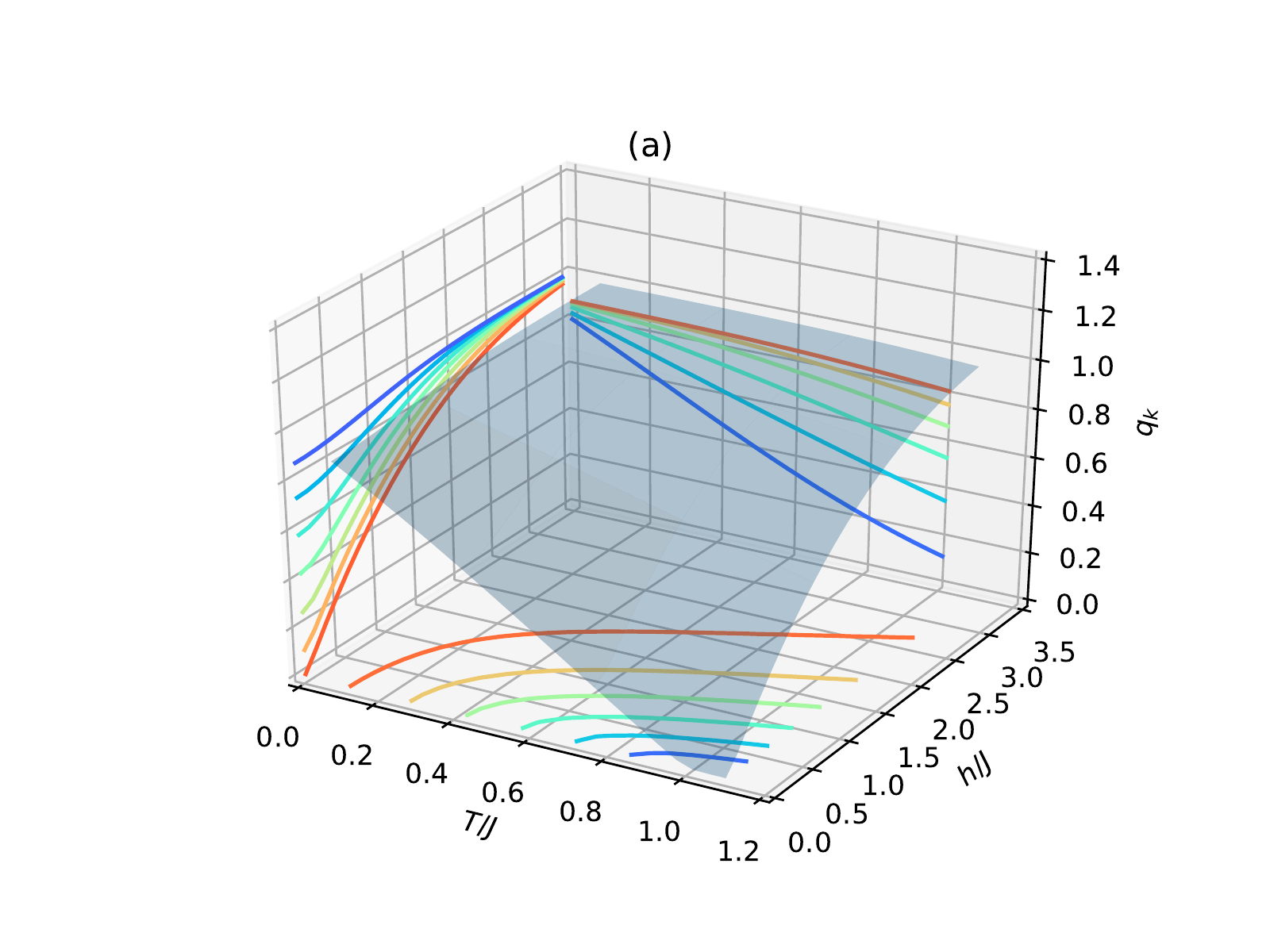} &
    \includegraphics[width=.34\textwidth]{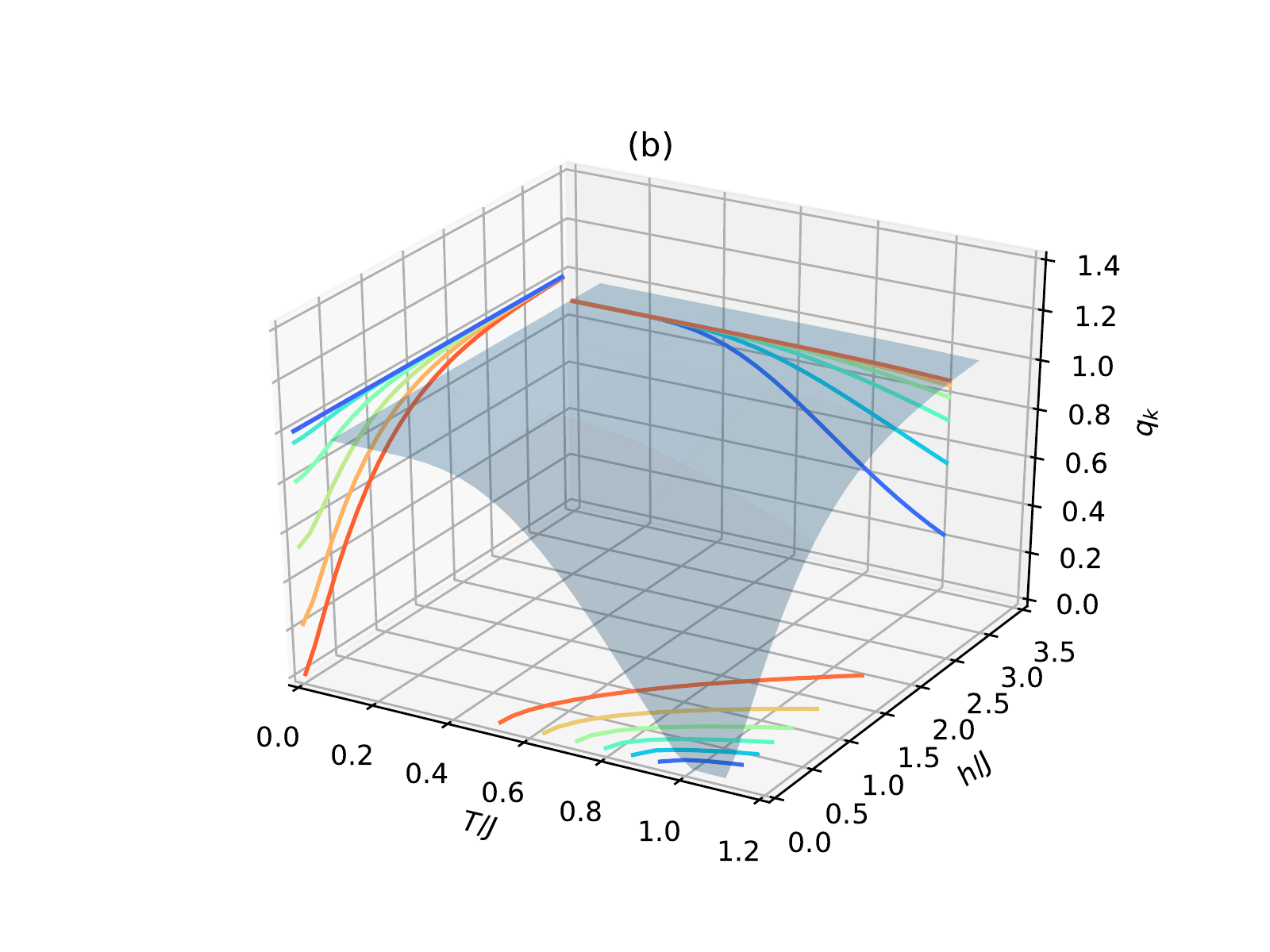} &
    \includegraphics[width=.34\textwidth]{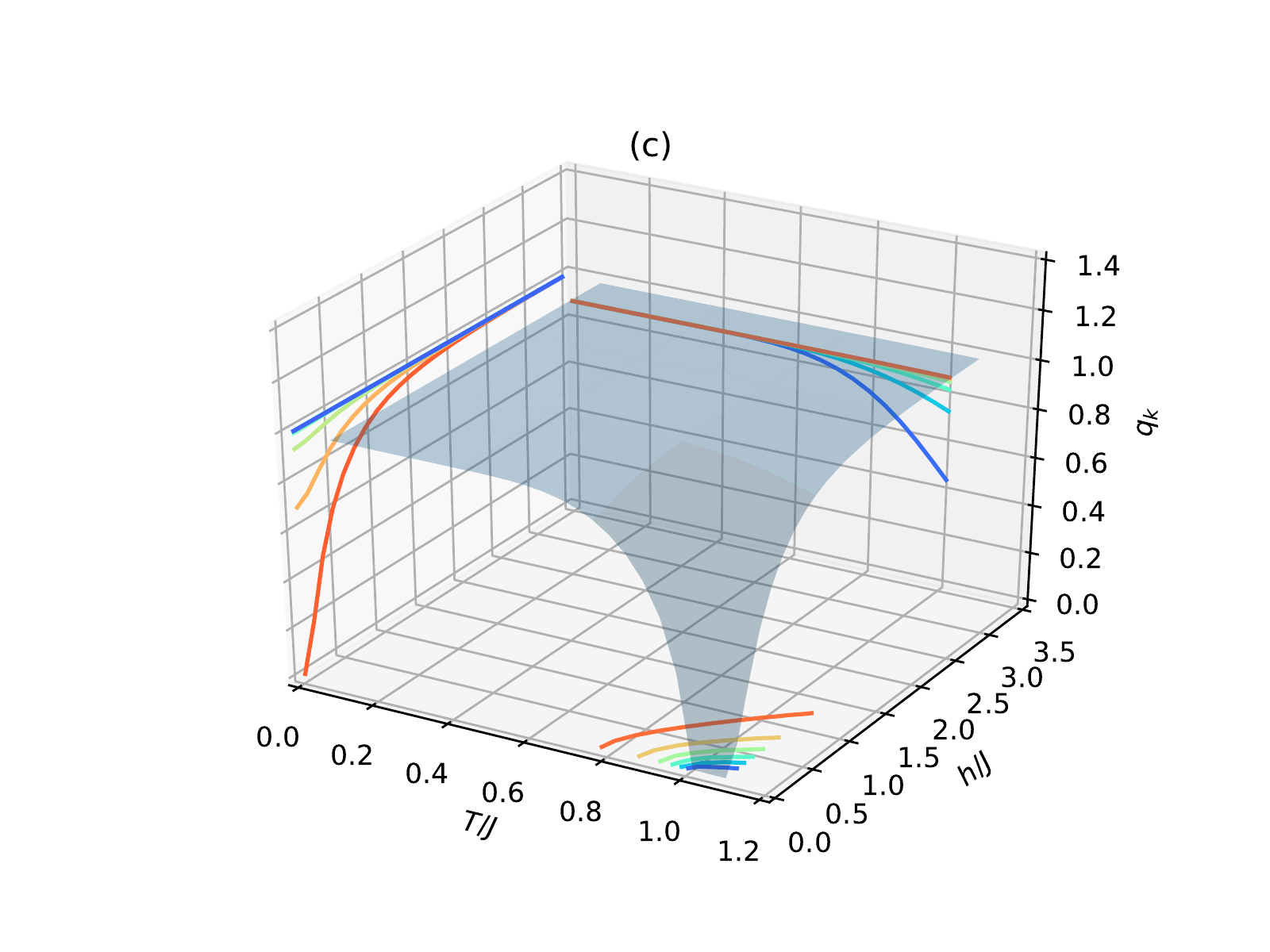} \\
    \includegraphics[width=.34\textwidth]{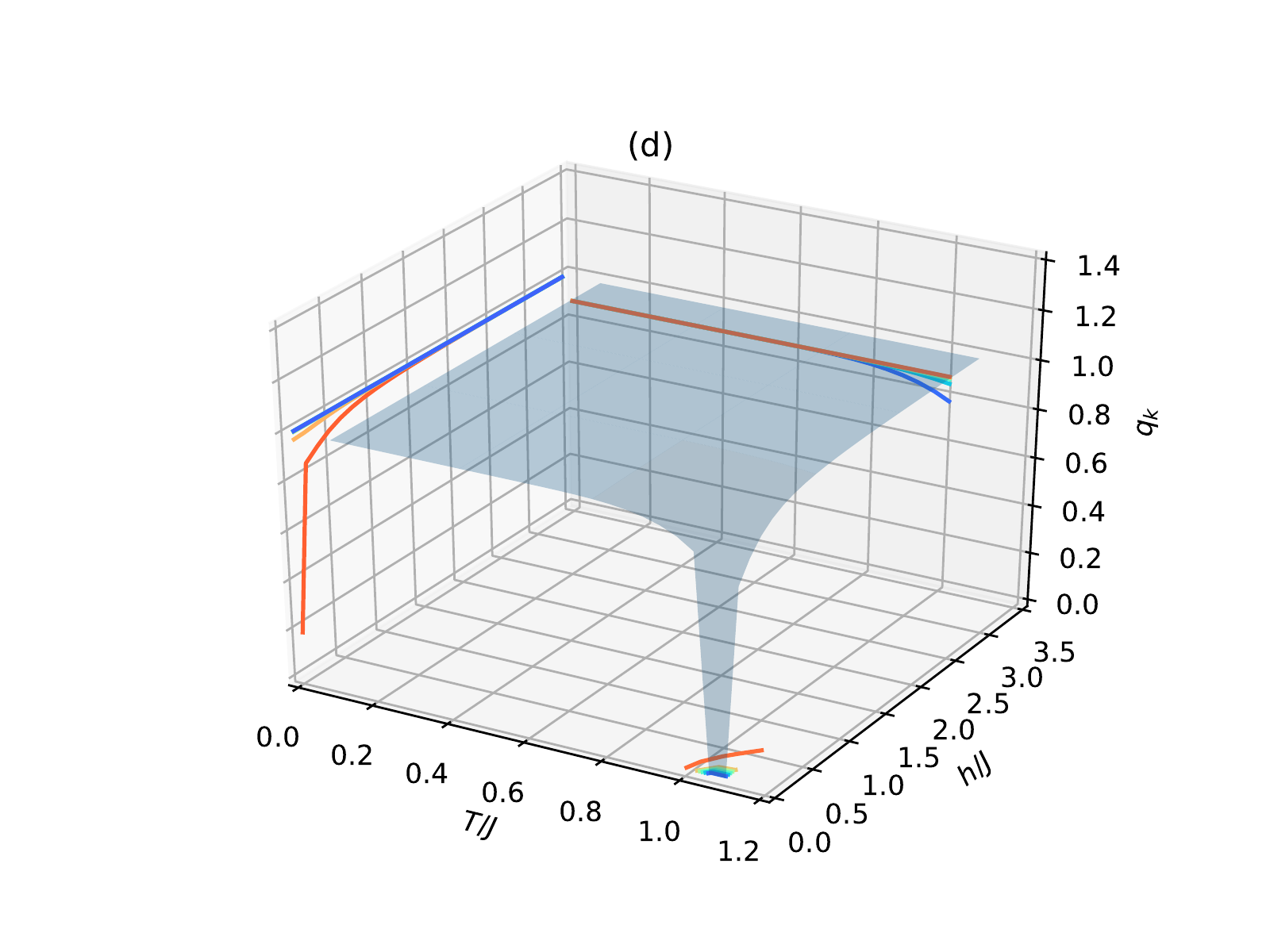} &
    \includegraphics[width=.34\textwidth]{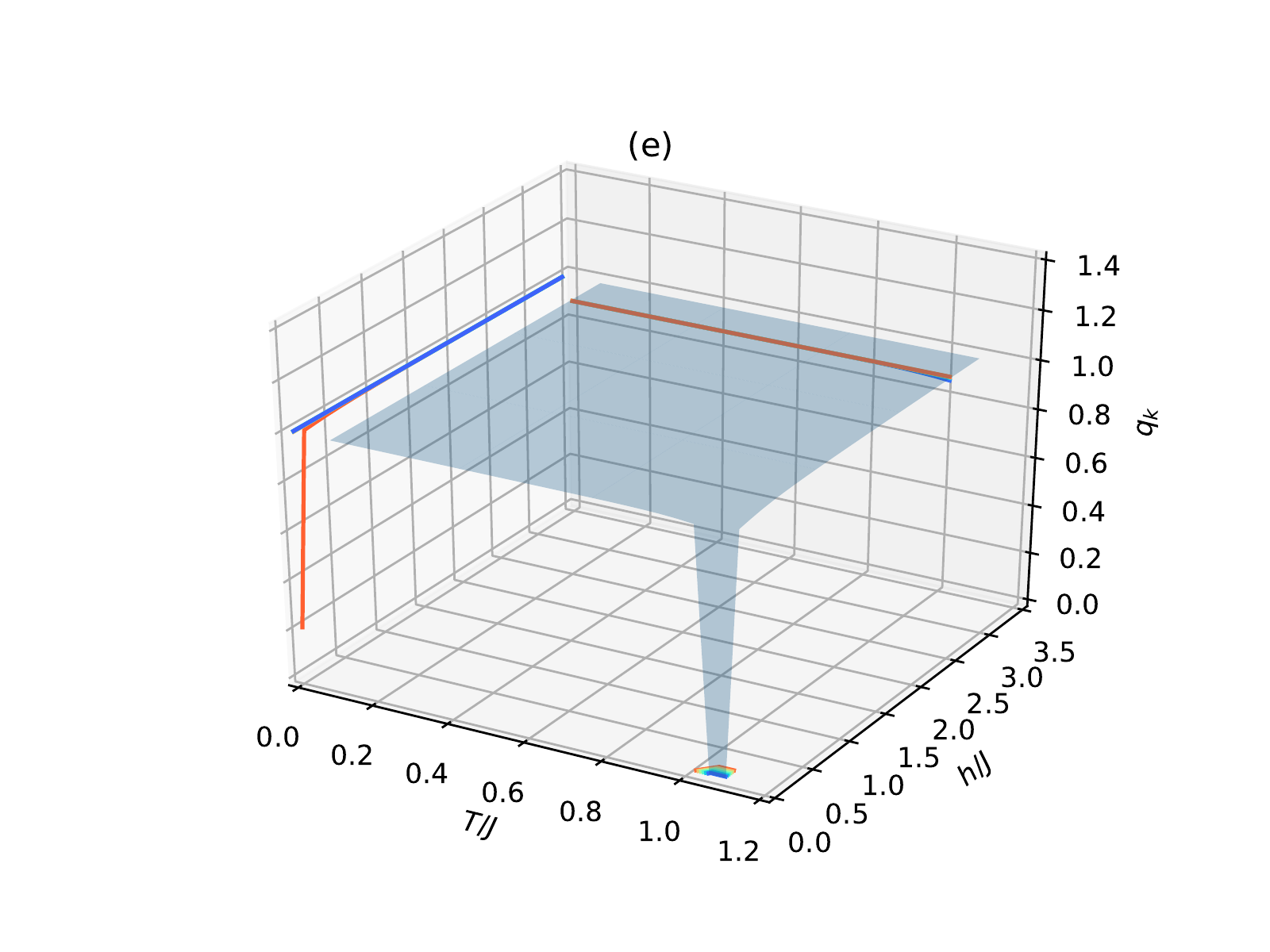} &
    \includegraphics[width=.34\textwidth]{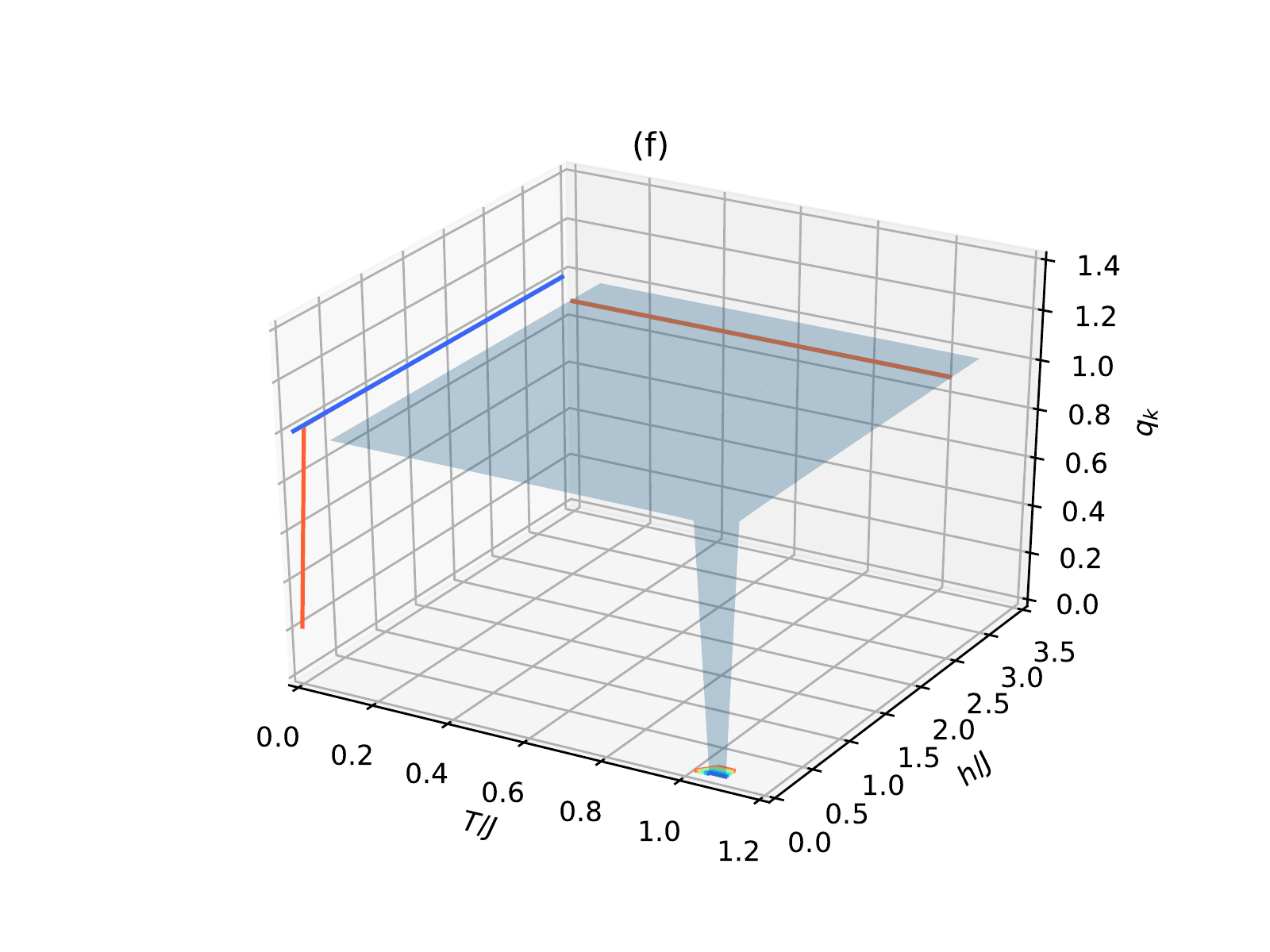} \\
  \end{tabular}
  \caption{Numerical solutions of (\ref{qzeta}) for (a) $k=0$ (replica symmetric ansatz), (b) $k=1$, (c) $k=2$, (d) $k=3$, (e) $k=4$, and (f) $k=5$. Here, the curves projected on each plane are level curves used to delimit the stability regions.}
  \label{qkk345}
\end{figure*}

Here, near the critical point the solution is stable if $1>\beta ^{2}J^{2}$. In the limit $T\rightarrow 0$ we have stability for all $k$. In FIG. (\ref{qkk345}) we depict numerical solutions of (\ref{qzeta}) when we can observe the behavior in the regions of stability, near the critical point, and for low fields and temperatures for $k=0,1, 2, 3, 4, 5$. Note that the discontinuity near the critical point is getting sharp as $k$ increases.

\section{Conclusions}
\label{conclusions}

Afterward the Parisi RSB scheme was consolidated as a solution to avoid unphysical scenarios obtained by the replica ansatz in the replica method, a wide set of results exploring each steep of the scheme, were discovering new physics of the Sherrington-Kirkpatrick model for spin glasses. Within this framework a great theoretical and numerical effort were carry out to study the physics of the complex free-energy landscape and order parameters under different domains. In this paper we adopt the distributional zeta function method (DZFM). There is a series representation of the average free energy where all of the moments of the partition function contribute.

Within the DZFM we obtain the multi-valley structure of the average free energy. We obtained the self-consistent integral equations for each order parameter $q_{k}$ and $m_{k}$, which have a similar structure presented in the replica symmetric ansatz and for the first steeps of the Parisi RSB. We perform a study of this parameter near the critical point and in the low temperatures regime. For the critical point we obtained a critical temperature for each $k$. Since we are dealing with a series representation, an asymptotic analysis was made to recover the known result $T_{C}=J$. In the evaluation of the local magnetizations and linear susceptibilities we have found similar behaviors described in phenomenological models and experimental results. In particular we obtain the behavior of reentrant phase in the local magnetization given by two critical temperatures determined by $J$ and $J_{0}$. On the other hand, with an upper order expansion, keeping terms of order $O(m_{k}^{2})$ and $O(q_{k}m_{k})$, we have obtained the typical discontinuity of the linear susceptibility at the spin-glass critical temperature. Furthermore, with numerical solutions of the integral equations we were able to recover the behavior of the continuum limit of the Parisi RSB. 

For the low temperature regime, we have shown that for each $k$ the limit result for $q_{k}$ is $1$ as has been investigated in the literature. Furthermore, before take the limit $\beta \rightarrow \infty$, we have obtained a composed expression for each $k$. The main result in this regime is the construction of a positive definite series representation of the ground state entropy. We show that the ground-state entropy goes to zero as temperature tends to zero, that is, in the low temperature regime, $\beta \rightarrow \infty$, we obtain $S\rightarrow 0$.

In the stability analysis of our results, we obtained a general expressions for the elements of the Hessian matrix of the free energy. We study the structure of the \emph{longitudinal} (scalar) space, which at the same time, can be one can be divided in other two orthogonal eigenspaces called, respectively, the \emph{anomalous} (vectorial) and \emph{replicon} (tensorial), for each $k$. From this structure we obtained our generalized version for the AT conditions for stability. Finally, we study the distribution of the overlaps. We obtained for each $k$ a structure similar (but not equal) to the ultrametricity extracted from the Parisi solution in the continuum limit. However, we have that as $k\rightarrow \infty$ the overlaps are becoming statistically independent. As we have shown, that distribution has not implications for the free-energy landscape because we are recovering its complexity thanks to the series expansion, i.e., we have a minimum within each term. In other words, we do not need to have states which are organized in a ultrametric fashion to obtain the multi-valley topography of the free energy. The expression (\ref{overlap111}), as is expected, is showing us that we do not have a ultrametric structure since we do not have a RSB that construct a hierarchy between the parameters extracted in each step~\cite{rammal1}. In addition, the obtained distribution is valid for $k>2$.

In conclusion, we have obtained analytical results that are able to access the complex free-energy landscape, its rich topography of metastable states, broken ergodicity, and multi-valley structure. Possible extensions of this work are the study of these effects over random graphs. For example, to give new insights over the finite size effects~\cite{ferrari1, lucibello1, metz1}, critical phenomena~\cite{mossi1, tikhonov1}, and random-link matching problems on random regular graphs~\cite{parisi12}. On the other hand we can use the DZFM on multiplex networks to investigate critical phenomena and collective behavior~\cite{bolfe1}, and finally use our formalism to enlarge the set of statistical field theory toolbox that is been currently used for simplicial complex~\cite{bianconi1, bianconi2}. These issues are under investigation by the authors.

\section*{Acknowledgments}

This paper was partially supported by the \emph{VIII Convocatoria para el Desarrollo y Fortalecimiento de los Grupos de Investigaci\'on en Uniminuto} with code C119-173 and Industrial Engineering Program from the Corporaci\'on Universitaria Minuto de Dios (Uniminuto, Colombia) (C.D.R.C.), and Conselho Nacional de Desenvolvimento Cient\'ifico e Tecnol\'ogico - CNPq, 303436/2015-8 (N.F.S.). We would like to thanks G. Krein and B. F. Svaiter for useful discussions about the mathematical and physical implications in whole the construction of this paper.
\section*{Appendix}
In this appendix we include further developments for the free energy from the DZFM and the overlap distribution under our formalism.
\appendix

\section{Replica method}
\label{replicamethodapp}

To study the properties and phase structure of the SK model it is necessary to compute the configurational averaged free energy defined by (\ref{averagefree111}). To bypass the average of the logarithm it is usually employed the replica method. After the construction of $Z^{k}$, the expected value of the partition function's $k$-th power $\mathbb{E}[Z^{k}(J_{ij})]$ is evaluated by integrating over the disorder field on the new model (collection of replicas). Notice that in $Z^{k}$, integration over quenched random couplings yields a system defined by $k$ replicas which are no more statistically independent. 

The average value in the presence of the quenched disorder is then obtained in the limit of a zero-component field theory taking the limit $k\rightarrow 0$ due to the following identity

\begin{equation}
\mathbb{E}[\log Z(J_{ij})]=\lim _{k\rightarrow 0}\frac{\mathbb{E}[Z^{k}(J_{ij})]-1}{k},
\label{replicam1}
\end{equation}
or
\begin{equation}
\mathbb{E}[\log Z(J_{ij})]= \lim _{k\rightarrow 0}\frac{\partial}{\partial k}\mathbb{E}[Z^{k}(J_{ij})] .
\end{equation}

The standard ansatz restricts the search for a minimum to a $q_{\alpha _{k}\gamma _{k}}$ invariant under a subgroup of the permutation group of $k$ elements. In the Parisi's RSB scheme, if $P_{k}$ is the group of permutations of $k$ elements, we can consider the chain
\begin{equation*}
P_{k}\supset (P_{m_{1}})^{k/m_{1}}\otimes P_{k/m_{1}},\, P_{m_{1}}\supset (P_{m_{2}})^{m_{1}/m_{2}}\otimes P_{m_{1}/m_{2}},
\end{equation*}
and so on. For instance, in the first step of replica symmetry breaking (1RSB), an invariant $6\times 6$ $q_{\alpha _{k}\gamma _{k}}$ is
\begin{equation*}
\begin{pmatrix}
0 & q_{1} & q_{1} & q_{0} & q_{0} & q_{0} \\
q_{1} & 0 & q_{1} & q_{0} & q_{0} & q_{0} \\
q_{1} & q_{1} & 0 & q_{0} & q_{0} & q_{0} \\
q_{0} & q_{0} & q_{0} & 0 & q_{1} & q_{1} \\
q_{0} & q_{0} & q_{0} & q_{1} & 0 & q_{1} \\
q_{0} & q_{0} & q_{0} & q_{1} & q_{1} & 0 
\end{pmatrix} .
\end{equation*}

While an $8\times 8$ $q_{\alpha _{k}\gamma _{k}}$  matrix invariant under $(P_{2})^{4}\otimes (P_{4/2})^{2}\otimes P_{8/4}$ is

\begin{equation*}
\begin{pmatrix}
0 & q_{0} & q_{1} & q_{1} & q_{2} & q_{2} & q_{2} & q_{2} \\
q_{0} & 0 & q_{1} & q_{1} & q_{2} & q_{2} & q_{2} & q_{2} \\
q_{1} & q_{1} & 0 & q_{0} & q_{2} & q_{2} & q_{2} & q_{2} \\
q_{1} & q_{1} & q_{0} & 0 & q_{2} & q_{2} & q_{2} & q_{2} \\
q_{2} & q_{2} & q_{2} & q_{2} & 0 & q_{0} & q_{1} & q_{1} \\
q_{2} & q_{2} & q_{2} & q_{2} & q_{0} & 0 & q_{1} & q_{1} \\ 
q_{2} & q_{2} & q_{2} & q_{2} & q_{1} & q_{1} & 0 & q_{0} \\
q_{2} & q_{2} & q_{2} & q_{2} & q_{1} & q_{1} & q_{0} & 0 
\end{pmatrix} .
\end{equation*}

The SK solutions correspond to using the replica-symmetric solution where $q_{\alpha _{k}\gamma _{k}}=q$ and $m_{\alpha _{k}}=m$, within the well known relation from the replica method (\ref{replicam1}), the order parameters $q$ and $m$ read, respectively,
\begin{equation}
q=1-\int Dz\, \text{sech}^{2}(\beta \psi) = \int Dz\, \text{tanh}^{2}(\beta \psi)
\label{qsymm1}
\end{equation}
and
\begin{equation}
m=\int Dz\, \text{tanh}(\beta \psi)
\label{msymm1}
\end{equation}
being
\begin{equation}
Dz = \frac{dz}{\sqrt{2\pi}}\exp \left(-\frac{z^{2}}{2}\right)
\end{equation}
and
\begin{equation}
\psi = J\sqrt{q}z+J_{0}m+h .
\end{equation}

In this scenario, we have the spin glass transition at $T=J$ when $J_{0}=m=h=0$. According to (\ref{qsymm1}) $q$ tends to one as $T\rightarrow 0$. However, the ground-state entropy is $-1/2\pi$. In order to avoid the unphysical results of the replica-symmetric solution, the replica symmetry breaking (RSB) was introduced. The variational parameters in the 1RSB are
\begin{equation}
m=\int Du\, \frac{\int Dv\, \text{cosh}^{m_{1}}\Xi \, \text{tanh}\Xi}{\int Dv\, \text{cosh}^{m1}\Xi},
\label{m1rsb}
\end{equation}
\begin{equation}
q_{0}=\int Du\, \left( \frac{\int Dv\, \text{cosh}^{m_{1}}\Xi \, \text{tanh}\Xi}{\int Dv\, \text{cosh}^{m1}\Xi} \right) ^{2},
\label{q01rsb}
\end{equation}
and
\begin{equation}
q_{1}=\int Du\, \frac{\int Dv\, \text{cosh}^{m_{1}}\Xi \, \text{tanh}^{2}\Xi}{\int Dv\, \text{cosh}^{m1}\Xi} ,
\label{q11rsb}
\end{equation}

being

\begin{equation*}
\Xi = \beta (J\sqrt{q_{0}}u+J\sqrt{q_{1}-q_{0}}v+J_{0}m+h) .
\end{equation*}

Under this formalism the entropy per spin at $J_{0}=0$, $T = 0$ reduces from $-0.16$ $(= -1/2\pi )$ for the RS solution to $-0.01$ for the 1RSB.

In the full RSB solution, since one obtains a $K\rightarrow \infty$ number of order parameters defined by
\begin{eqnarray}
\sum _{\alpha _{k}\neq\gamma _{k}}q_{\alpha _{k}\gamma _{k}}^{l} = q_{0}^{l}k^{2}+(q_{1}^{l}-q_{0}^{l})m_{1}^{2}\frac{k}{m_{1}}
\nonumber\\
+ (q_{2}^{l}-q_{1}^{l})m_{2}^{2}\frac{m_{1}}{m_{2}}\frac{k}{m_{1}}+\cdots - q_{K}^{l}k .
\end{eqnarray}

Rewriting the last expression we get
\begin{equation}
\sum _{\alpha _{k}\neq\gamma _{k}}q_{\alpha _{k}\gamma _{k}}^{l} = k\sum _{j=0}^{K}(m_{j}-m_{j+1})q_{j}^{l},
\end{equation}
where $l$ is an arbitrary integer and $m_{0}=k$, $m_{K+1}01$. In the limit $k\rightarrow 0$, we may use the replacement $m_{j}-m_{j+1}\rightarrow -dx$ to find

\begin{equation}
\frac{1}{k}\sum _{\alpha _{k}\neq \beta _{k}}q_{\alpha _{k} \beta _{k}}^{l} \rightarrow -\int _{0}^{1}q^{l}(x)dx .
\end{equation}

Near the critical point this parameter has the following behavior

\begin{equation}
q(x)=\begin{cases}
\frac{x}{2}\quad \text{if}\quad 0\leq x \leq x_{1}=2q(1)\\
q(1) \quad \text{if}\quad x_{1}\leq x \leq 1 
\end{cases} ,
\label{parisifullrsb}
\end{equation}
being
\begin{equation}
q(1)=| \theta | +O(\theta ^{2}), \quad \theta \ll 1.
\end{equation}

\section{Free energy from DZFM}
\label{dzfmapp1}
In this appendix we review the alternative approach to compute the configurational average of the free energy of disordered systems presented in~\cite{svaiter1, svaiter2, svaiter3}. We begin with the definition of the generalized $\zeta$-function given by
\begin{equation}
\zeta _{\mu ,f}(s)=\int _{X}f(x)^{-s}d\mu (x),
\end{equation}
where the triplet $(X,\mathcal{A},\mu)$ is a measure space, $f:X\rightarrow (0,\infty)$ is measurable, and $s\in \mathbb{C}$ such that $f^{-s}\in L^{1}(\mu)$, where in the above integral $f^{-s}=\exp (-s\log (f))$ is obtained using the principal branch of the logarithm. In the situation where $f(J_{ij})=Z(J_{ij})$ and $d\mu (J_{ij})$ we obtain the definition of the distributional zeta function $\Phi (s)$ as
\begin{equation}
\Phi (s)=\int d[J_{ij}]P(J_{ij})\frac{1}{Z(J_{ij})^{s}} .
\label{phiapp1}
\end{equation} 

Following the usual steps of the spectral zeta function, the configurational average free energy can be written as
\begin{equation}
f=\lim _{N\rightarrow \infty}\frac{1}{N\beta}\left.\frac{d}{ds}\Phi (s)\right| _{s=0^{+}},\quad \text{Re}(s)\geq 0 ,
\end{equation}
where $\Phi (s)$ is well defined. The procedure use the Euler's integral representation of the gamma function
\begin{equation}
\frac{1}{Z(J_{ij})^{s}} = \frac{1}{\Gamma (s)}\int _{0}^{\infty} dt\, t^{s-1}\text{e}^{-Z(J_{ij})t},
\label{gammarep1}
\end{equation}
for $\text{Re}(s)\geq 0$. Substituting (\ref{gammarep1}) into (\ref{phiapp1}), we have
\begin{equation}
\Phi (s)=\frac{1}{\Gamma (s)}\int d[J_{ij}]P(J_{ij})\int _{0}^{\infty} dt\, t^{s-1}\text{e}^{-Z(J_{ij})t} .
\end{equation}

To proceed, it is assumed the commutativity of the configurational average, differentiation and integration. Now, we take $a>0$ and write $\Phi = \Phi _{1}+\Phi _{2}$ where
\begin{equation}
\Phi _{1}(s)=\frac{1}{\Gamma (s)}\int d[J_{ij}]P(J_{ij})\int _{0}^{a} dt\, t^{s-1}\text{e}^{-Z(J_{ij})t}
\end{equation}
and
\begin{equation}
\Phi _{2}(s)=\frac{1}{\Gamma (s)}\int d[J_{ij}]P(J_{ij})\int _{a}^{\infty} dt\, t^{s-1}\text{e}^{-Z(J_{ij})t},
\end{equation}
being $a$ a dimensionless parameter. The configurational average free energy can be written as
\begin{equation}
f = \lim _{N\rightarrow \infty}\frac{1}{N\beta} \left\lbrace \left.\frac{d}{ds}\Phi _{1} (s)\right| _{s=0^{+}} +\left.\frac{d}{ds}\Phi _{2}(s)\right| _{s=0^{+}} \right\rbrace .
\end{equation}

In the innermost integral in $\Phi _{1}$, the series representation for the exponential converges uniformly (for each $J_{ij}$), so that we can reverse the order of integration and summation to obtain
\begin{equation}
\Phi _{1}(s)=\int d[J_{ij}]P(J_{ij})\frac{1}{\Gamma (s)} \sum _{k=0}^{\infty}\frac{(-1)^{k}a^{k+s}}{k!(k+s)}Z^{k}(J_{ij}) .
\label{phi1appint}
\end{equation} 

The term $k=0$ in (\ref{phi1appint}) contains a removable singularity at $s=0$ since $s\Gamma (s) = \Gamma (s+1)$, so that we can write
\begin{equation}
\Phi _{1}(s)=\frac{a^{s}}{\Gamma (s+1)}+\frac{1}{\Gamma (s)}\sum _{k=0}^{\infty}\frac{(-1)^{k}a^{k+s}}{k!(k+s)}\mathbb{E}[ Z^{k}] .
\end{equation}

The function $\Gamma (s)$ has a pole at $s=0$ with residue 1, therefore
\begin{equation}
-\left.\frac{d}{ds}\Phi _{1} (s)\right| _{s=0^{+}} = \sum _{k=0}^{\infty}\frac{(-1)^{k+1}a^{k}}{k!k}\mathbb{E}[ Z^{k}] +f(a) ,
\end{equation}
where 
\begin{equation}
f(a)=-\left. \frac{d}{ds}\left( \frac{a^{s}}{\Gamma (s+1)}\right)\right| _{s=0^{+}} = -\log (a)+\gamma _{e}
\end{equation}
being $\gamma _{e}$ the Euler's constant. The derivative of $\Phi _{2}$ is given by
\begin{equation}
\left.\frac{d}{ds}\Phi _{2} (s)\right| _{s=0^{+}} = \int d[J_{ij}]P(J_{ij}) \int _{a}^{\infty}\frac{dt}{t}\, \text{e}^{-Z(J_{ij})t}=R(a) .
\end{equation}

The asymptotic behavior of $R(a)$ is related to the incomplete gamma function defined as
\begin{equation}
\Gamma (\alpha ,x)=\int _{x}^{\infty} \text{e}^{-t}t^{\alpha -1}\, dt .
\end{equation}
The asymptotic representation for $|x|\rightarrow \infty$ and $-3\pi /2 < \text{arg}x<3\pi /2$ reads
\begin{equation}
\Gamma (\alpha ,x)\sim x^{\alpha -1}\text{e}^{-x}\left[ 1+\frac{\alpha -1}{x}+\frac{(\alpha -1)(\alpha -2)}{x^{2}}+...\right] .
\end{equation}

Defining $Z_{c}$ as the partition function of a system where $P(J_{ij})=c$, where $c\in \mathbb{R}$ is a constant such that $c<J_{ij}$ for all stochastic variable $J_{ij}$ defined by (\ref{distributionj}), we have a bound for $R(a)$ given by $|R(a)|\leq (Z_{c}a)^{-1}\exp (-Z_{c}a)$.

\section{Overlap distribution}
\label{overlapapp1}

For completeness in our analysis we present the overlap distribution under our formalism. In the Parisi RSB this distribution is associated with the continuum parameter $x$ of the order parameter $q(x)$. Furthermore, the joint distribution shows that the symmetry breaking generates a ultrametric structure. The explicit calculation of $P(Q_{1},Q_{2},Q_{3})$ gives the result
\begin{eqnarray}
P(Q_{1},Q_{2},Q_{3})= &\frac{1}{2}&P(Q_{1})x(Q_{1})\delta (Q_{1}-Q_{2})\delta (Q_{2}-Q_{3})
\nonumber \\
&+&\frac{1}{2}P(Q_{1})P(Q_{2})\Theta (Q_{1}-Q_{2})\delta (Q_{2}-Q_{3})
\nonumber \\
&+&\frac{1}{2}P(Q_{2})P(Q_{3})\Theta (Q_{2}-Q_{3})\delta (Q_{3}-Q_{1})
\nonumber \\
&+&\frac{1}{2}P(Q_{3})P(Q_{1})\Theta (Q_{3}-Q_{1})\delta (Q_{1}-Q_{2}) .
\nonumber \\
\end{eqnarray} 

For any $Q_{c}<Q_{max}$ the integrated contributions in the volume $0\leq Q_{i}\leq Q_{C}$ of all four terms are equal. It follows that, among all the triangles, $1/4$ are equilateral and $3/4$ isosceles. The relatively large number of equilateral triangles is a noticeable feature of this kind of ultrametricity. Now let us show what happens with this quantity under the DZFM.

The thermodynamic average of the site magnetization could be represented as
\begin{equation}
\left\langle \sigma _{i} \right\rangle = m_{i}=\sum _{\alpha _{k}}w_{\alpha _{k}}m_{i}^{\alpha _{k}},
\end{equation}
where the $\alpha _{k}$'s label the pure states and $w_{\alpha _{k}}$ are their statistical weights which could be written as
\begin{equation}
w_{\alpha _{k}}=\text{e}^{-F_{\alpha _{k}}},
\end{equation}
being $F_{\alpha _{k}}$ the free energy of the pure state $\alpha$.

The overlap between two states $\alpha _{k}$ and $\gamma _{k}$ is defined by
\begin{equation}
Q_{\alpha _{k} \gamma _{k}}=\frac{1}{N}\sum _{i=1}^{N}m_{i}^{\alpha _{k}}m_{i}^{\gamma _{k}} .
\end{equation}

We can observe that $0\leq |Q_{\alpha _{k} \gamma _{k}}|\leq 1$. To describe the statistics of the overlaps between all the pairs of pure states it is natural to introduce the probability distribution function
\begin{equation}
P_{J}(Q)=\sum _{\alpha _{k} \gamma _{k}}w_{\alpha _{k}}w_{\gamma _{k}}\delta (Q_{\alpha _{k} \gamma _{k}}-Q) .
\end{equation}

The function $P_{J}(Q)$ could depend on the concrete realization of the quenched interactions $J_{ij}$. The average over the disorder is $P(Q)=\mathbb{E}[P_{J}(Q)]$. The function $P(Q)$ gives the probability to find two pure states having mutual overlap equal to $Q$. In terms of the entries of the matrix $q_{\alpha _{k} \gamma _{k}}$ we have
\begin{equation}
P(Q)=\frac{2}{k(k-1)}\sum _{\alpha _{k} < \gamma _{k}}\delta (q_{\alpha _{k} \gamma _{k}}-Q) .
\end{equation}

To explore the metric of the space of pure states we consider the distribution function $P(Q_{1},Q_{2},Q_{3})$ which describes the joint statistics of the overlaps of arbitrary three pure states. We have then
\begin{widetext}
\begin{eqnarray}
P(Q_{1},Q_{2},Q_{3})=\frac{1}{k(k-1)(k-2)} \sum _{\alpha _{k} \neq \gamma _{k} \neq \sigma _{k}} \delta (q_{\alpha _{k} \gamma _{k}}-Q_{1})\delta (q_{\alpha _{k} \sigma _{k}}-Q_{2})\delta (q_{\gamma _{k} \sigma _{k}}-Q_{3}) .
\end{eqnarray}

To perform the calculus, we use the Fourier transform of the function $P(Q_{1},Q_{2},Q_{3})$
\begin{equation}
g(y_{1},y_{2},y_{3})=\int dQ_{1}dQ_{2}dQ_{3}\, P(Q_{1},Q_{2},Q_{3})\, \text{e}^{iQ_{1}y_{1}+iQ_{2}y_{2}+iQ_{3}y_{3}} .
\end{equation}

Thus we have
\begin{eqnarray}
g(y_{1},y_{2},y_{3})&=&\frac{1}{k(k-1)(k-2)}\sum _{\alpha _{k} \neq \gamma _{k} \neq \sigma _{k}}\text{e}^{iq_{\alpha _{k} \sigma _{k}}y_{1}+iq_{\alpha _{k} \sigma _{k}}y_{2}+iq_{\gamma _{k} \sigma _{k}}y_{3}}
\nonumber\\
&=&\frac{1}{k(k-1)(k-2)}\text{Tr}[A(y_{1})A(y_{2})A(y_{3})]
\label{g123}
\end{eqnarray}

where
\begin{equation}
A_{\alpha _{k} \gamma _{k}}(y)=\begin{cases}
\text{e}^{iq_{\alpha _{k} \gamma _{k}}y}\quad \text{if}\quad \alpha _{k} \neq \gamma _{k} \\
0 \quad \text{if}\quad \alpha _{k} = \gamma _{k}
\end{cases} .
\end{equation}

Applying the fact that $q_{\alpha _{k} \gamma _{k}}\rightarrow q_{k}$ for each term in the free energy, the trace in (\ref{g123}) yields
\begin{eqnarray}
\text{Tr}[A(y_{1})A(y_{2})A(y_{3})] = k\left\lbrace 1+ (k-1)\left[ \text{e}^{iq_{k}(y_{1}+y_{2})}+\text{e}^{iq_{k}(y_{1}+y_{3})}+\text{e}^{iq_{k}(y_{2}+y_{3})} \right]+(k-1)(k-2)\text{e}^{iq_{k}(y_{1}+y_{2}+y_{3})} \right\rbrace .
\end{eqnarray} 

Finally we obtain
\begin{eqnarray}
P_{k}(Q_{1},Q_{2},Q_{3})=& &\frac{1}{8}P(Q_{1})P(Q_{2})P(Q_{3})+\frac{1}{4(k-2)}\left[P(Q_{1})P(Q_{2})\delta (Q_{3})+P(Q_{1})P(Q_{3})\delta (Q_{2})+P(Q_{2})P(Q_{3})\delta (Q_{1})\right]
\nonumber \\
&+&\frac{1}{(k-1)(k-2)}\delta (Q_{1})\delta (Q_{2})\delta (Q_{3}) .
\label{overlap111}
\end{eqnarray}

\end{widetext}

As is expected, we do not have a ultrametric structure since we do not have a RSB that construct a hierarchy between the parameters extracted in each step~\cite{rammal1}. Note that as $k\rightarrow \infty$ the overlaps are becoming statistically independent.

\end{document}